\documentclass[aps,pra,twocolumn,amsfonts,superscriptaddress,showpacs]{revtex4-1}

\usepackage{amsmath,amssymb,mathrsfs}
\usepackage{latexsym}
\usepackage{graphicx} 
\usepackage{epstopdf}
\usepackage{graphicx,epstopdf,color}
\usepackage{amsfonts}
\usepackage{amsmath,amssymb,mathrsfs}
\usepackage{subfigure}
\usepackage{hyperref}
\usepackage{bm}

\newcommand{\ket}[1]{\left|#1\right\rangle}
\newcommand{\bra}[1]{\left\langle#1\right|}

\def\eg{\emph{e.g.}\ }

\allowdisplaybreaks[1]
\begin{document}
\title{Anomalous Hall Effects of Light and Chiral Edge Modes on the Kagome Lattice}
\author{Alexandru Petrescu}
\affiliation{Department of Physics, Yale University, New Haven, CT 06520, USA}
\author{Andrew A. Houck}
\affiliation{Department of Electrical Engineering, Princeton University, Princeton, New Jersey 08544, USA}
\author{Karyn Le Hur}
\affiliation{Centre de Physique Th\' eorique, \' Ecole Polytechnique, CNRS, 91128 Palaiseau C\' edex, France}
\affiliation{Department of Physics, Yale University, New Haven, CT 06520, USA}
\date{\today}

\begin{abstract}
We theoretically investigate a photonic Kagome lattice which can be realized in microwave cavity arrays using current technology. The Kagome lattice exhibits an exotic band structure with three bands one of which can be made completely flat. The presence of artificial gauge fields allows to emulate topological phases and induce chiral edge modes which can coexist inside the energy gap with the flat band that is topologically trivial. By tuning the artificial fluxes or in the presence of disorder, the flat band can also acquire a bandwidth in energy allowing the coexistence between chiral edge modes and bulk extended states; in this case the chiral modes become fragile towards scattering into the bulk. The photonic system then exhibits equivalents of both a quantum Hall effect without Landau levels, and an anomalous Hall effect characterized by a non-quantized Chern number. We discuss experimental observables such as local density of states and edge currents. We show how synthetic uniform magnetic fields can be engineered, which allows an experimental probe of Landau levels in the photonic Kagome lattice. We then draw on semiclassical Boltzmann equations for transport to devise a method to measure Berry's phases around loops in the Brillouin zone. The method is based solely on wavepacket interference and can be used to determine band Chern numbers or the photonic equivalent of the anomalous Hall response. We demonstrate the robustness of these measurements towards on-site and gauge-field disorder. We also show the stability of the anomalous quantum Hall phase for nonlinear cavities and for (artificial) atom-photon interactions.
\end{abstract}

\maketitle
\section{Introduction}
\begin{figure}[t!]
\includegraphics[width=0.90\linewidth]{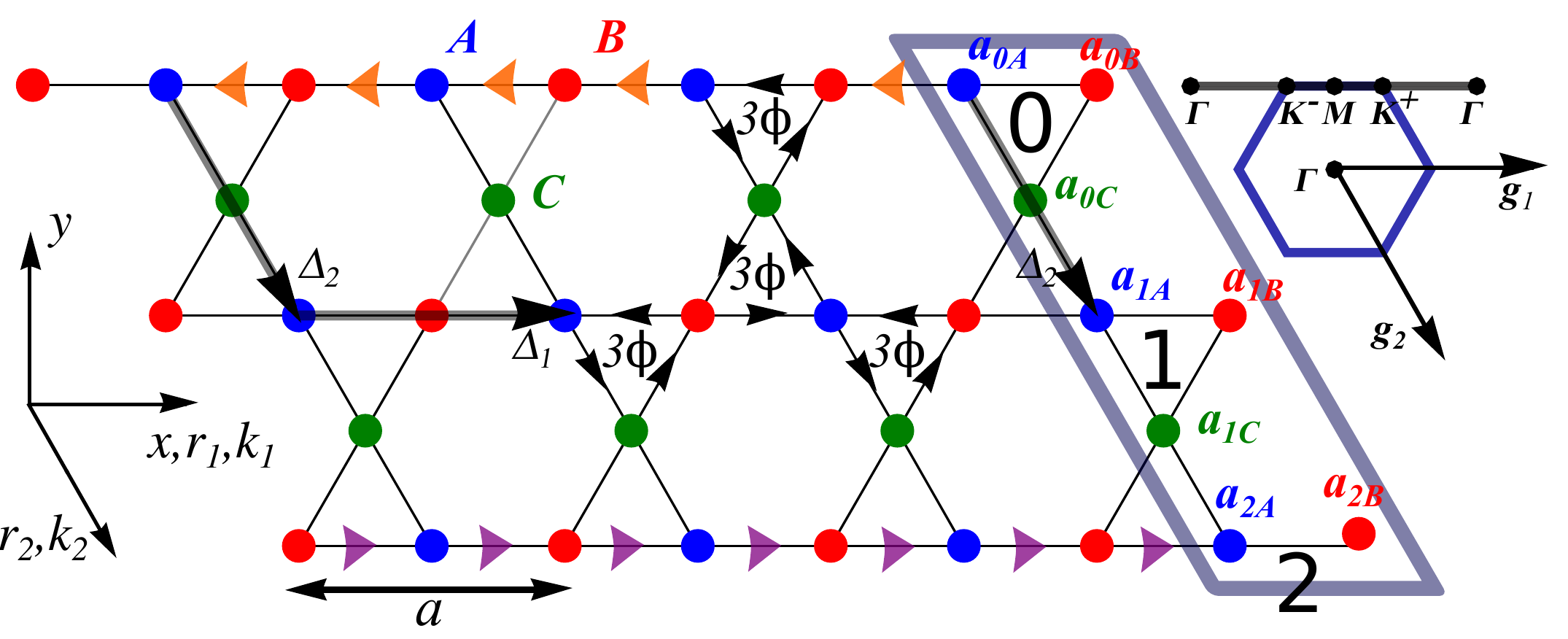}
\includegraphics[width=0.45\linewidth]{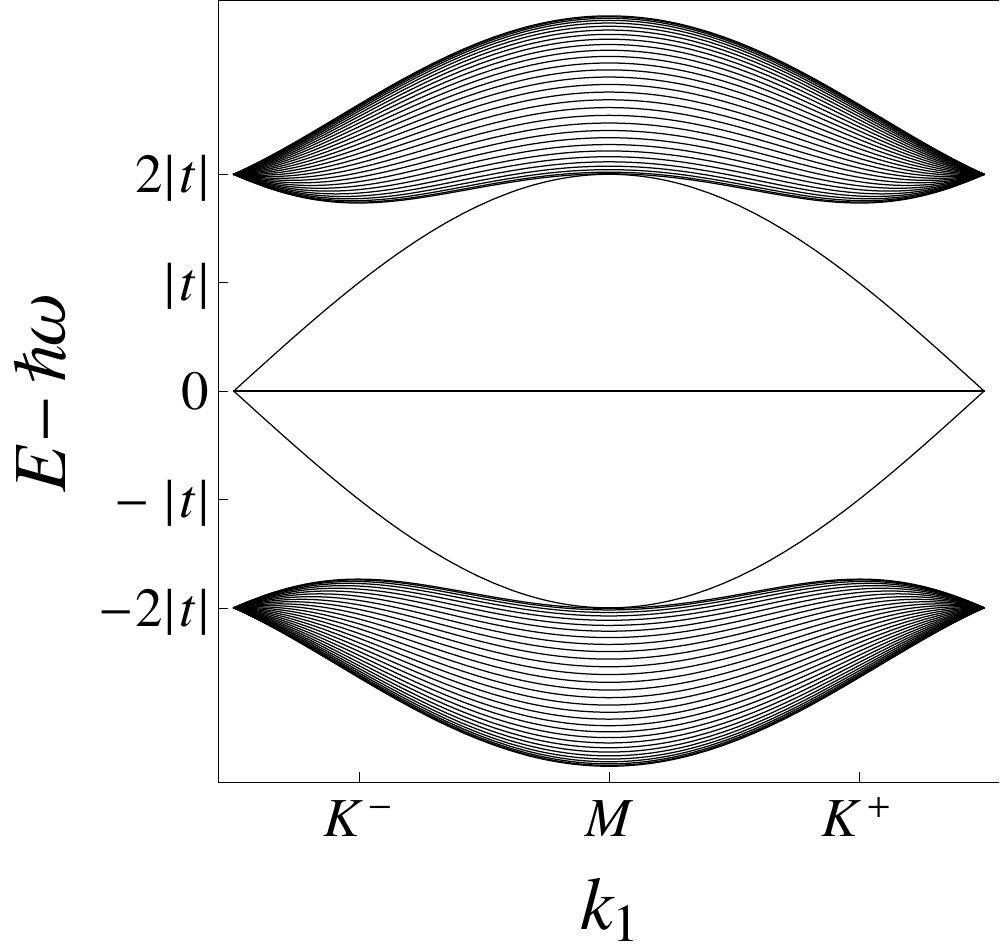}
\includegraphics[width=0.45\linewidth]{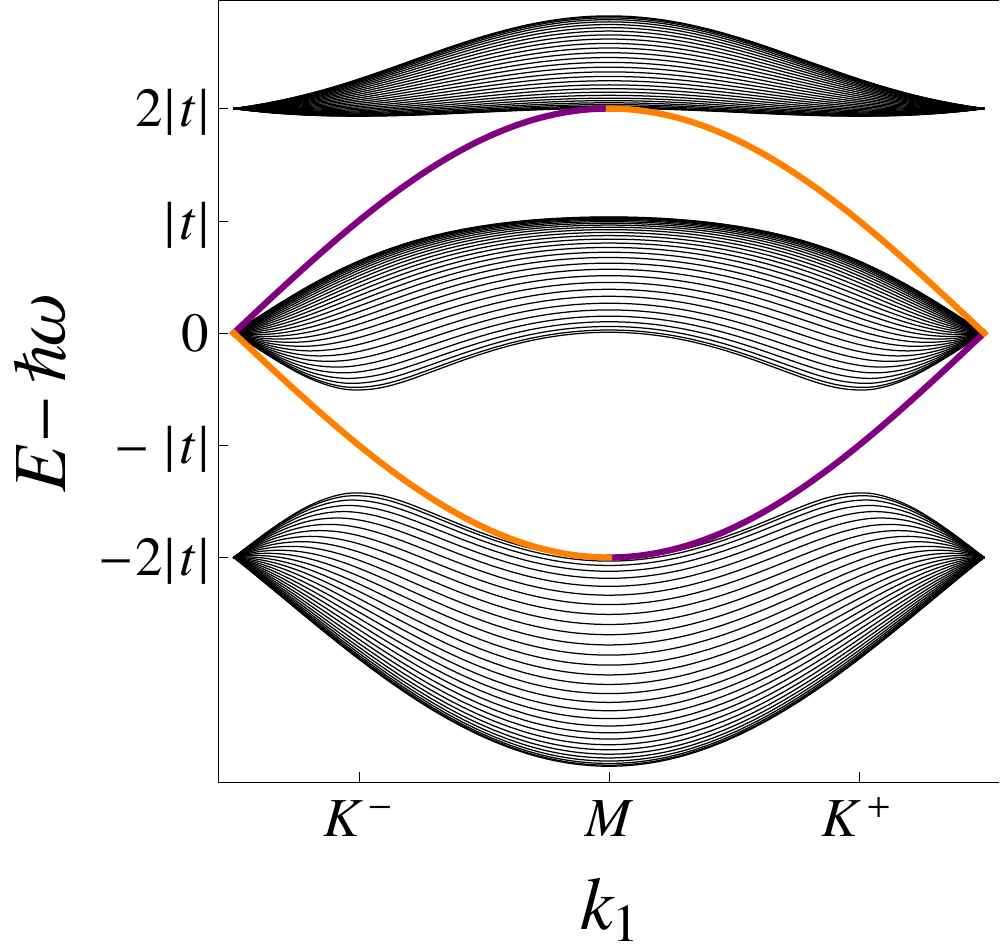}
\caption{\label{fig:k_BS}
(Color online) The photonic Kagome lattice with three sites per unit cell $A,B,C$. Artificial fluxes can be threaded through the triangular plaquettes \cite{Kagpho}. Top left: lattice periodic in the $x$ direction, with ``line'' boundaries composed of $A,B$ sites only. The blue parallelogram encloses a one-dimensional superunit cell which generates the entire lattice. Top right: a projection of the Brillouin zone onto the $k_x\equiv k_1$ direction. Bottom left: at $\phi=\pi/6$ the system exhibits a flat, topologically trivial middle band, while the lower and upper bands are characterized by finite Chern numbers. Bottom right: the middle band becomes dispersive if one deviates from $\pi/6$, \eg for $\phi=\pi/4$. Here, we have considered a cylinder geometry which allows for two chiral edge modes: right-moving at the lower edge (purple), left-moving at the upper edge (orange); corresponding chiral currents depicted on the lattice. As discussed in Secs. IV B and C, the chiral edge modes can also be detected in the square geometry \cite{Houck}.}
\end{figure}

Phenomena associated with flat bands in fermionic and bosonic systems have drawn attention over the last three decades \cite{Mielke,Wu,Green,Wen,Mudry,Sarma,Regnault,Cooper}. The best known example is perhaps the fractional quantum Hall effect \cite{Laughlin}. If a single-particle band is dispersionless in one direction, electrons are localized in that direction causing an element of the effective mass tensor to diverge. If the band is completely flat in the entire $k$-space, then heavy degeneracy appears and the density of states diverges. Such singularities in the density of states are expected to produce anomalous behaviors in physical properties including transport phenomena and optical response. If a wavepacket is created in the flat band system, the group velocity will automatically vanish and very strong backscattering prevents the packet from moving. In fact, this scenario takes place on the Kagome lattice where localized states on hexagon rings occur as a result of the alternating sign of the wavefunction amplitude \cite{Tasaki,Nishino}.

The topological properties of the band structure are now essential in approaching the problem of Chern insulators \cite{Haldane1988} or topological insulators \cite{Kane, Zhang}. Bandstructure topology has been first discussed in the context of the integer quantum hall effect by Thouless {\it et al.} \cite{Thouless}.

More precisely, the set of energy eigenstates that form an isolated band is described by the first Chern number \cite{Chernf}, a topological invariant associated with the band. The Chern number has direct physical consequences: for example, a completely filled (electronic) band has a quantized Hall conductance, corresponding to the existence of chiral edge modes \cite{Hatsugai}. It is possible to view this invariant as the flux of the Berry curvature \cite{Berry} through the first Brillouin zone \cite{Chern}. The Berry curvature and the Chern number have become important mathematical concepts to classify topological phases in connection with band structures. The Chern number has previously been accessed through the conductance and chiral edge states of quantum Hall systems, and the total Berry phase associated with a single Dirac point has been inferred from transport in graphene \cite{experiments}. Measuring the local Berry curvature constitutes an actual challenge in experiments. 

In this paper, we investigate anomalous Hall phases of light on a specific Kagome lattice with artificial magnetic fluxes, which has been introduced in the context of photonic lattices \cite{Kagpho}; see Fig. \ref{fig:k_BS}. The effect of time-reversal symmetry breaking in connection with this Kagome photon lattice has been previously studied at the level of a few site geometry \cite{Kagpho}.  We show below that this lattice with artificial gauge fields allows for the existence of an anomalous quantum  Hall phase, or a quantum Hall phase without Landau levels \cite{Haldane1988}. 

Photonic lattices based on arrays of circuit quantum electrodynamics (cQED) superconducting elements \cite{SCircuit} have been recently realized experimentally \cite{Houck,HouckTureciKoch}. Such photon cavity lattices are predicted to exhibit interesting many-body phenomena \cite{PlenioReview2008,FazioReview} including a superfluid-Mott transition of light \cite{JensKarynJC,KLH,Greentree,Bose,Plenio,RossiniFazio, Aichhorn,NaEtAl, SchmidtBlatter},  Bose-Hubbard models with attractive interactions \cite{LeibHartmann}, fractional quantum Hall physics \cite{Sougato,Greentree2,UmucalilarCarusottoFQH} and interesting dynamics \cite{Hakanetal} in particular in ring geometries \cite{Andreas}. Recently a Rabi model has been shown to present a $Z_2$ Ising universality class quantum phase transition between two gapped phases, which are not Mott insulator phases \cite{SchiroTureci}.  An anomalous quantum Hall phase for light accompanied by unidirectional photonic chiral edge states has been predicted in photonic crystals \cite{HaldaneRaghu,Wang} and confirmed experimentally \cite{MIT}.  In addition, two-dimensional photonic equivalents of topological insulators have been envisaged \cite{Demler,MacDonald}. Topological properties of optical systems might help to implement robust photonic devices \cite{Demler} and to realize invisibility cloaks \cite{Pendry}. Artificial gauge fields in photonic systems have attracted growing attention \cite{CarusottoCiuti, UmucalilarCarusotto}. 

Our primary goal is to investigate the effect of the topologically trivial flat band (with a zero Chern number) on the robustness of photonic chiral edge modes. In our realization of the Kagome lattice, neighboring hexagons are subject to the same magnetic flux. The magnetic flux opens a gap at the Dirac points and breaks time-reversal symmetry allowing to stabilize topological phases characterized by two bands acquiring non-zero Chern numbers (see Fig. \ref{fig:k_BS}). The middle band can be made flat or acquire a finite energy dispersion for certain values of the artificial gauge fields (see Fig. \ref{fig:k_BS}) or if disorder is present; therefore, edge modes can coexist with extended bulk states at the same energy. In this situation edge modes are fragile towards leaking into the bulk resulting in an anomalous Hall phase with a non-quantized Chern number \cite{Haldane2004}. A similar effect has occurred in a toy model on the honeycomb lattice \cite{Doron}. We investigate the evolution of unidirectional edge modes of light in the presence of a localized scatterer by tuning the artificial gauge fields and by introducing disorder. We present a manner to determine Berry's phases of photon wavepackets through an interference experiment. There exists a previous proposal to access the Berry curvature through a measurement of group velocities of wavepackets in optical lattices \cite{PriceCooper}. We discuss a direct measurement of the Chern number of a Bloch band, which does not require summation of contributions across the Brillouin zone. Underlying these methods is the implementation of uniform synthetic gauge fields through tunable cavities with time-dependent frequencies, and the realization of Landau levels in photon systems. It shall be noticed that a similar scheme has been recently realized in cold atom systems \cite{StruckEtAl} and that cavities with tunable resonances have been already implemented in cQED \cite{Sandberg,DelsingPRL}. 

In addition, the quest for exotic phases on the Kagome lattice represents an active subject of research 
\cite{Lhuillier, Lhuillier2,Balents,Guo,Santos,Isakov,Wen,Fiete,Melko,Balents2, HuseWhite}. Topological phases and their supporting artificial gauge fields are potentially realizable in cold atom systems \cite{Spielman,Dalibard,JakschZoller, Jimenez-GarciaEtAl, StruckEtAl2, AidelsburgerEtAl}. The Hofstadter spectrum \cite{Hofstadter} on the related dice lattice has been realized in GaAlAs/GaAs systems \cite{NaudEtAl}. The Hofstadter spectrum for the Kagome lattice has also been studied with superconducting wire networks \cite{XiaoEtAl}. Finally, the Kagome structure naturally appears in real materials \cite{Mendels,Lhuiller3} and recently it has been implemented in cold atom systems \cite{Berkeley}.

The paper is organized as follows. We introduce the tight-binding model of a Kagome lattice in the presence of artificial gauge fields in Sec. \ref{k_tbmodel}. Explicit solutions for the edge mode wavefunctions in cylinder geometry and a brief discussion of the bulk-edge correspondence are the subjects of Sec. \ref{k_edgemodes}. Following this, in Sec. \ref{k_obs} we present observables specific to the anomalous Hall phase, namely the bulk polarization, the edge currents, and the local density of states. Sec. \ref{k_berryphaseobs} further details on a method to measure Berry's phases of wavepackets in both clean and disordered systems. The method requires a uniform synthetic magnetic field, on whose realization in a photonic lattice we elaborate. Additionally, we show a method to directly find the topological Chern number of a given Bloch band by counting energy levels of the system in an artificial magnetic field. In Sec. \ref{k_chern}, we discuss in detail the anomalous Hall effects of light in relation to Chern numbers. In the particular case of disordered systems, we present a real-space calculation of the Chern number which will reveal a close analogy between tuning the artificial gauge fields and the presence of disorder in the system (in cQED lattices, disorder can appear either as an on-site scalar or as a vector potential). In Sec. \ref{k_sym} we discuss the role of symmetries and the local stability of Dirac points in the three band system. In Sec. \ref{k_int} we study the effect of interactions in QED cavities \cite{RaimondBruneHaroche}, such as Bose-Hubbard \cite{nonlinear} and Jaynes-Cummings \cite{JC}, on the topological phase. The Appendices are dedicated to technical details.

\section{Time-reversal Symmetry Breaking}
\label{k_tbmodel}
Below, we introduce the Kagome lattice with artificial gauge fields. An effective tight-binding Hamiltonian for photons with the possibility to break time-reversal symmetry can be realized in a cQED system by coupling superconducting waveguides to nano-Josephson circulators (rings) \cite{Kagpho}. The Kagome lattice is formed by triangular and hexagonal plaquettes arranged as shown in Fig. \ref{fig:k_BS}. The resulting structure has three sites per unit cell, which we denote by $A,B,C$. Letting $a$ be twice the length of a bond, we pick the following lattice vectors $\mathbf{\Delta}_{1,2}$ and reciprocal lattice vectors $\mathbf{g}_{1,2}$:
\begin{eqnarray}
{\bf\Delta}_1=a(1,0), &\;&\; {\bf \Delta}_2=a\left(\frac{1}{2},-\frac{\sqrt{3}}{2}\right) \nonumber \\
{\bf g}_1=\frac{4\pi}{\sqrt{3}a}\left(1,0\right), &\;\;& {\bf g}_2=\frac{4\pi}{\sqrt{3}a}\left(\frac{1}{2},-\frac{\sqrt{3}}{2}\right).
\end{eqnarray}
The Brillouin zone, see Fig. \ref{fig:k_BS}, is hexagonal and contains the common points of high symmetry $\Gamma=(0,0)$ and $K^\pm =\pm\left(\frac{4\pi}{3a},0\right)$. 

We will use the following lattice coordinates: the directions along $\hat{\mathbf{\Delta}}_1$, $\hat{\mathbf{\Delta}}_2$ in real-space corresponding to $\hat{\mathbf{g}}_1, \hat{\mathbf{g}}_2$ in momentum space (top right panel of Fig. \ref{fig:k_BS}), where hats denote unit vectors. In this notation, $\hat{\Delta}_1$ coincides with the $x$ direction, $\hat{x}$. We denote sites on the lattice by a pair of integers $\mathbf{m}=(m_1,m_2)$, and coordinates on the lattice in terms of lattice vectors by $\mathbf{r}_\mathbf{m}\equiv m_1 \mathbf{\Delta}_1 + m_2\mathbf{\Delta}_2$. Continuous coordinates are represented by a pair of real numbers $\mathbf{r}=(r_1, r_2)$ in terms of unit vectors, $\mathbf{r}=r_1 \hat{\mathbf{\Delta}}_1+r_2 \hat{\mathbf{\Delta}}_2$; momenta will be written as $\mathbf{k}=k_1 \hat{\mathbf{g}}_1 + k_2 \hat{\mathbf{g}}_2$.

In a cQED system, a given waveguide of length in the millimeter range typically supports a single photonic mode. Furthermore, the photons which travel in these one-dimensional waveguides do not carry a polarization label and must be thought of as simply excitations of microwave resonators \cite{DevoretReview}. The hopping of photons from one waveguide to another results in a translationally invariant tight-binding Hamiltonian. Following Ref. \cite{Kagpho}, a honeycomb array of waveguides with nano-Josephson circulators can be equivalently reformulated as the photonic Kagome lattice of Fig. \ref{fig:k_BS}. The Hamiltonian can be Fourier transformed, and we obtain:
\begin{equation}
\label{eq:k_H}
\mathscr{H}=\sum_{{\bf k} \in \text{BZ}}\psi_{{\bf k}}^{\dagger}\mathscr{H}_{{\bf k}}\psi_{{\bf k}},
\end{equation}
\noindent and the spinor $\psi_{{\bf k}}^{\dagger}=\left(a_{A{\bf k}}^{\dagger},a_{B{\bf k}}^{\dagger},a_{C{\bf k}}^{\dagger}\right)$ contains the creation operators on each sublattice, and

\begin{equation}
\label{eq:k_Hk}
\mathscr{H}_{\mathbf{k}} = 
\left(\begin{array}{ccc}
\hbar\omega & 2|t| e^{i\phi}\cos \alpha_1 & 2|t| e^{-i\phi}\cos \alpha_{2}\\
2|t| e^{-i\phi}\cos \alpha_1 & \hbar\omega & 2|t| e^{i\phi}\cos \alpha_{12}\\
2|t| e^{i\phi}\cos \alpha_{2} & 2|t| e^{-i\phi}\cos \alpha_{12} & \hbar\omega
\end{array}\right).
\end{equation}
\noindent We have defined three dimensionless functions of momentum,
\begin{equation}
\alpha_1(\mathbf{k}) \equiv \mathbf{k}.\frac{\mathbf{\Delta}_1}{2} ,\; \alpha_2(\mathbf{k}) \equiv \mathbf{k}.\frac{\mathbf{\Delta}_2}{2} ,\; \alpha_{12}(\mathbf{k}) \equiv  \mathbf{k}.\frac{\mathbf{\Delta}_1-\mathbf{\Delta}_2}{2}. 
\end{equation}

In the effective tight-binding model particles hop between nearest-neighbor sites with a complex hopping integral $|t| e^{\pm i \phi}$. Photons acquire a phase $3\phi$ around a triangular plaquette and a phase of $-6 \phi$ around the hexagonal plaquette, amounting to zero total flux in the parallelogram unit cell of area $|\mathbf{\Delta}_1 \times \mathbf{\Delta}_2|$, which is characteristic of the anomalous quantum Hall effect without Landau levels as introduced by Haldane \cite{Haldane1988}.  The particles are also subject to an artificial on-site potential induced by the (electromagnetic superconducting resonator) waveguide frequency $\omega$, and an artifical gauge field giving rise to complex hopping integrals and breaking time-reversal symmetry. It shall be noted that both the effective on-site potential $\hbar\omega$ and the hopping strength $|t|$ should be obtained rigorously after integrating out coupling element degrees of freedom (in our case, a Josephson ring coupling three resonators) \cite{Kagpho}. In a typical cQED experiment, the energy of an incoming photon is much larger than the size of the energy gap between bands: $|t|$ is expected to lie below $100$ MHz and the frequency $\omega$ is in the GHz, or microwave, range. 

The rotational symmetry of the lattice is not strictly necessary and anisotropies are possible; their effect on the bandstructure is discussed in Sec. \ref{k_sym}. In the rest of the paper we shall focus on the isotropic case.

The Hamiltonian of Eq. \ref{eq:k_Hk} is a band Hamiltonian obeying Bloch's theorem. To determine the Bloch energies $E(\mathbf{k})$ for the three bands let us denote $U(\mathbf{k})\equiv-(E(\mathbf{k})-\hbar\omega)/|t|$, and $\alpha(\mathbf{k})\equiv\cos\alpha_1(\mathbf{k})\cos\alpha_2(\mathbf{k})\cos(\alpha_{12}(\mathbf{k}))$. We obtain the following equation for the eigenvalues
\begin{equation}
\label{eq:k_evals}
U^{3}(\mathbf{k})-4U(\mathbf{k})(2\alpha(\mathbf{k})+1)+16\cos(3\phi)\alpha(\mathbf{k})=0.
\end{equation}
For special values of the phase $\phi$, destructive interference confines the wavefunction to the hexagonal plaquettes, which renders the respective band completely dispersionless. We obtain such a solution with energy $U(\mathbf{k})=2\cos3\phi$ whenever $\cos(3\phi)(\cos3\phi-1)(\cos3\phi+1)=0$. The middle band is flat whenever $\phi = \frac{\pi}{6} + \frac{m\pi}{3}$, where $m$ is an integer. When 
$\phi = \frac{m\pi}{3}$, the upper or lower band is flat, touching the middle band at the $\Gamma$ point; the other two bands form Dirac cones at the six corners of the Brillouin zone. For example, the bandstructure for the cases $\phi = \frac{\pi}{6}$ and $\frac{\pi}{4}$ (which 
will be thoroughly discussed in the text) is depicted in Fig. \ref{fig:k_BS}, along with edge modes arising from a finite geometry to be discussed below. The edge modes appear because by tuning the phase $\phi$ one breaks time-reversal symmetry at the Dirac points. The bandstructure of Eq. (\ref{eq:k_evals}) is invariant to a shift of $\frac{2\pi}{3}$ in $\phi$; we then restrict to $\phi \in \left[ 0, \frac{2\pi}{3}\right]$ in the further discussion.

\begin{figure}[t!]
\includegraphics[width=0.82\linewidth]{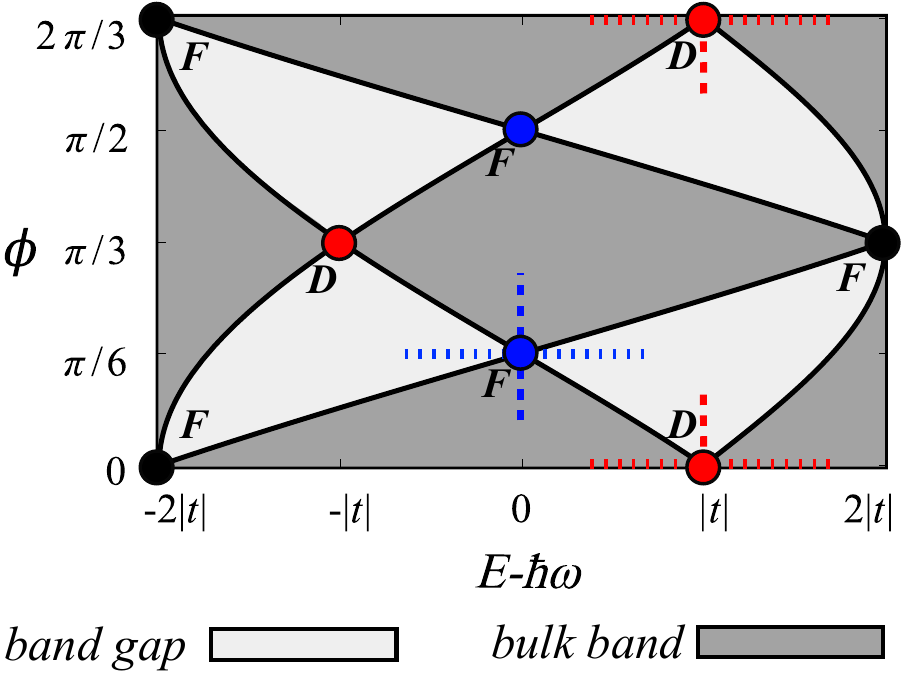}
\caption{(Color online) Two-dimensional phase diagram of a particle (photon) with energy $E$ on the Kagome lattice defined with hopping phase $\phi$: by tuning $(E,\phi)$ the photon system is either characterized by extended bulk states (``bulk band'')  or by a ``band gap'' where the system exhibits unidirectional edge modes but no bulk states. The \textit{F} points represent states in the flat band (the blue points at $E-\hbar\omega = 0$ correspond to a middle flat band separated by finite energy gaps from the lower and upper bands; black points at $E-\hbar \omega = \pm 2 |t|$ correspond to flat upper or lower band). The \textit{D} points are degeneracy points between pairs of bands at the Dirac points $K^\pm$. The dashed lines are explained in the main text. \label{fig:k_phases}}
\end{figure}

Time-reversal symmetry is kept if the phases accumulated along any closed loop on the lattice amount to an integer multiple of $\pi$ \cite{Kagpho}. This condition is equivalent to requiring that $\mathscr{H}_\mathbf{-k}$ and its time-reversed counterpart, $\mathscr{H}_\mathbf{k}^*$, are the same up to a gauge transformation on the spinor $\psi_\mathbf{k}$ in Eq. (\ref{eq:k_H}). In our case, if $3\phi$ is an integer multiple of $\pi$, then the Hamiltonian is time-reversal symmetric. The bandstructure properties can be summarized in the $(E,\phi)$ phase diagram of  Fig. \ref{fig:k_phases}. The phase diagram represented is periodic in $\phi$. A particle created at energies in the ``band gap'' regions cannot be in a two-dimensional Bloch state of the Hamiltonian. States exist in this energy interval only in a finite geometry with boundaries, as we show in detail in Sec. \ref{k_edgemodes}. Such a state would be confined to the edges of the sample and propagate chirally (chiral edge mode). The ``bulk band'' phase is embodied by bulk extended states. When photonic bulk extended states and chiral edge modes coexist at the same energy $E$, the edge excitations may scatter into the bulk; this situation coincides with an anomalous Hall phase. The \textit{F} points in the diagram of Fig. \ref{fig:k_phases} correspond to flat bands. Dashed lines from the $F$ points show the possible phase transitions: horizontal blue line - from the flat band to a band gap by changing the particle energy; vertical blue line: from the flat band to a bulk band by changing the phase $\phi$. The \textit{D} points represent band degeneracies at the Dirac points $K^\pm$. Moving along the vertical red line (a change in $\phi$) lifts the degeneracy and a finite band gap is opened. A particle with a slightly larger energy would be in a bulk band state.

At $\phi = 0$, the system is gapless: there is a quadratic touching with a flat band at the $\Gamma$ point, corresponding to $F$ in Fig. \ref{fig:k_phases}, and Dirac cones at the $K$ points, corresponding to $D$. We will be focusing on the system in the vicinity of $\phi = \frac{\pi}{6}$, at which the system exhibits a flat band. Changing the energy of the particle at the $F$ points (equivalent to moving along a horizontal line in the figure) will take us from the flat band into a band gap. Changing $\phi$ at the $F$ point (equivalent to moving along the vertical line) will take us from the flat band into a bulk-band with finite bandwidth.

\section{Chiral edge modes}
\label{k_edgemodes}
In this section we make explicit the correspondence between a bulk quantity, the Chern number, and the chiral edge modes at the edges of a sample. It should be noted that even though we choose the cylindrical  geometry below for mathematical convenience, the edge states can be detected in a square geometry; this point will be illustrated below in Figs. 5 and 6. 

Given a translationally invariant two-dimensional lattice Hamiltonian whose eigenstates are the Bloch states, the existence of chiral edge modes is signaled by the non-zero values of the Chern number $\nu^{(n)}$ corresponding to the Bloch band $|n \mathbf{k}\rangle$ \cite{Thouless},
\begin{equation}
\label{eq:k_chernkspace}
\nu^{(n)} = \frac{1}{2\pi}\int_{\text{BZ}} d^2\mathbf{k} \left(\partial_{\mathbf{k}} \times \mathscr{R}^{(n)}({\mathbf{k}}) \right),
\end{equation}
\noindent where the vector field $\mathscr{R}^{(n)}(\mathbf{k})$ is the Berry gauge potential associated to the $n^{th}$ Bloch band,
\begin{equation}
\label{eq:k_BerryGauge}
\mathscr{R}^{(n)}(\mathbf{k}) = -i \langle n \mathbf{k} | \partial_\mathbf{k} |n \mathbf{k} \rangle.
\end{equation}
The integral in Eq. (\ref{eq:k_chernkspace}) is over the entire surface of the Brillouin zone, so it is a summation over all of the states enclosed in the $n^{th}$ band. 

When the bands are separated by an energy gap throughout the Brillouin zone, the band Chern numbers of the lower, middle, and upper bands, respectively, can be shown numerically to be $-\text{sgn}\left(\sin{3\phi}\right), 0$ and $+\text{sgn}\left(\sin{3\phi}\right)$, where $\phi \in \left[ 0,\frac{2\pi}{3}\right]$. The $\sin(3\phi)$ function was chosen to obey the $2\pi/3$ periodicity of the bandstructure with respect to the phase $\phi$. In particular, this is consistent with a recent result that implies that the flat band of the Kagome lattice, if isolated from the other bands by energy gaps (as is the case for $\phi = \frac{\pi}{6}$), will be non-topological \cite{Katsura}.

The Chern number of Eq. (\ref{eq:k_chernkspace}) is a topological index whose value can only change if by variation of parameters the bands touch at some degeneracy points in the Brillouin zone and subsequently reopen the energy gap \cite{Thouless}. As the parameter $\phi$ is varied, for values  $0, \frac{\pi}{3},  \frac{2\pi}{3}...$ two bands touch at all corners of the Brillouin zone forming Dirac points; the remaining pair has a degeneracy at the $\Gamma$ point. At the transition upper and lower bands exchange Chern numbers leaving the middle band topologically trivial throughout. The stability of Dirac points and the nature of the exchange of Chern numbers is explained in Sec. \ref{k_sym}.  We now make the correspondence between the bulk quantity and the edge modes explicit. Consider an analogous fermion system at zero temperature: all states with energies lower than the chemical potential $\mu$ are occupied. If $\mu$ is set to an energy gap, the \text{sum} of band Chern numbers for the occupied bands is an integer quantity, possibly zero, which is proportional to the Hall conductivity \cite{Thouless}. The principle of bulk-edge correspondence states that this sum counts the chiral edge modes \cite{Hatsugai} supported at the given chemical potential. Since the argument is independent of statistics, it applies equally well for photons; we will then replace the notion of ``chemical potential'' by ``energy of injected particle''.

\begin{figure}[t!]
\centering
\includegraphics[width=\linewidth]{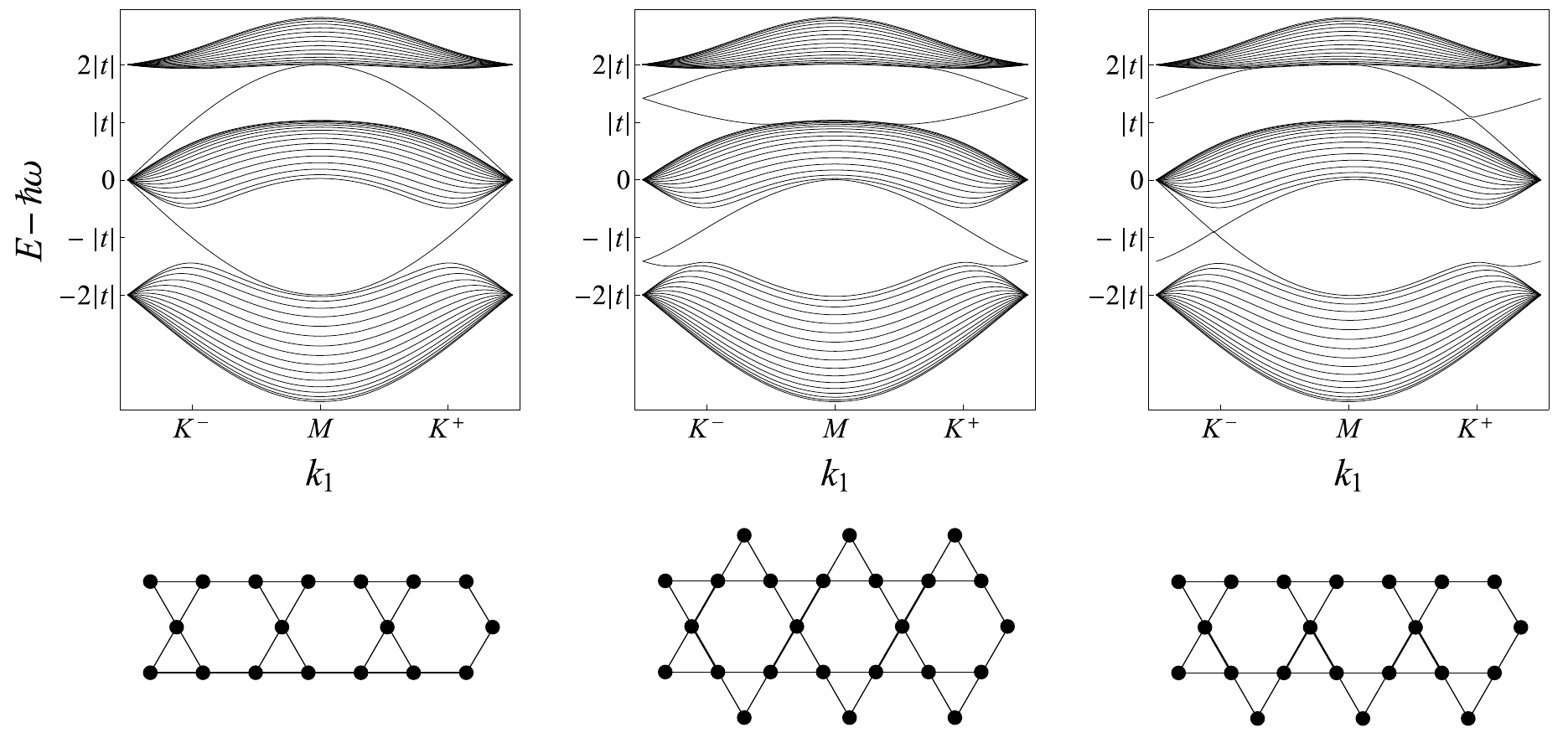}
\caption{\label{fig:k_bc} Effect of different boundary conditions on the dispersion of edge modes in the proximity of the dispersive middle band for a system with $\phi = \frac{\pi}{4}$: left, two ``line'' boundaries; middle, one ``line'' edge, the other ``armchair''; right, both boundaries are ``armchair''. The bottom panels exhibit the minimal ladder lattices that exhibit the three boundary conditions. All three ladders shown support edge modes. }
\end{figure}

\begin{figure}[t!]
\includegraphics[width=0.6\linewidth]{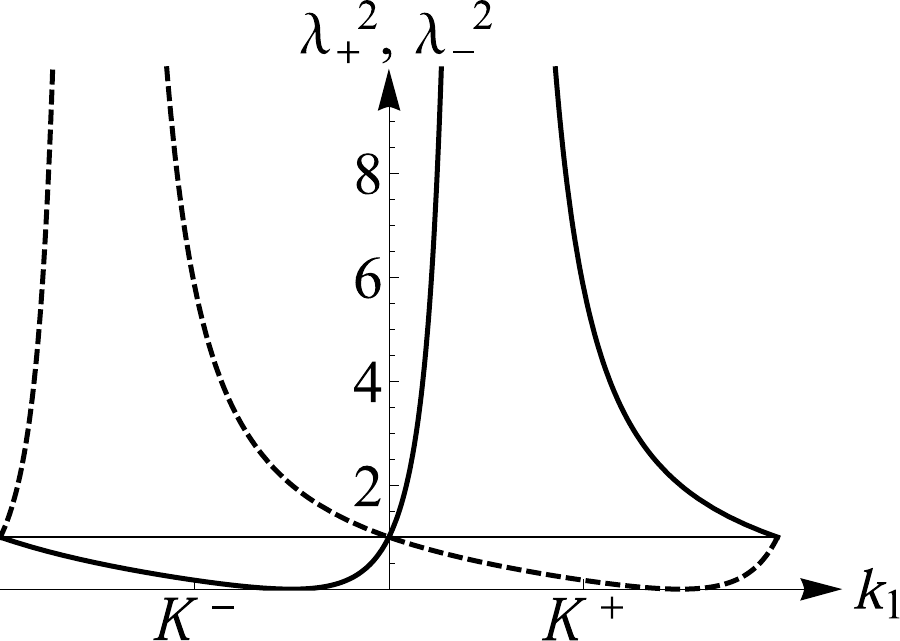}
\caption{\label{fig:k_lambdas} The quantities $\lambda_+^2$ (solid line), and $\lambda_-^2$ (dashed line) defined in Eqs. (\ref{eq:k_edgeStateEnFn}) describe how the wavefunction of the edge mode decays into the bulk.The reference line is at $\lambda_\pm^2=1$. For  $|\lambda_\pm|>1$  the respective branch $E_\pm(k_1)$ is located at the bottom edge. Conversely, if $|\lambda_\pm|<1$, it is located at the top edge. This results in the color scheme for the edge modes on Fig. \ref{fig:k_BS}}.
\end{figure}

Let us place the Kagome lattice in a cylinder geometry: a particle injected at an energy inside the energy gap can belong to one of two states localized at the edges. We now find an analytical solution of the edge wavefunctions. We consider a cylinder with line boundaries along the periodic $\hat{\mathbf{\Delta}}_1$ direction (boundaries containing $A,B$ sites only - see Fig. \ref{fig:k_bc}). The cylinder is generated by translating the superunit cell (blue parallelogram in Fig. \ref{fig:k_BS}) along the lattice vector $\mathbf{\Delta}_1$. The superunit cell is obtained by translating the triangular plaquette along $\mathbf{\Delta}_{2}$. We number plaquettes along the superunit cell by integer $m_2 = 0, 1,..., m_{2\text{max}}$, as in Fig. \ref{fig:k_BS}. We Fourier transform the operators $a_{\mathbf{m}\equiv(m_1,m_2)}$ along the periodic $k_1$ direction
\begin{equation}
a_{\alpha m_2}(k_1) = \frac{1}{\sqrt{N_1}} \sum_{m_1} e^{-i k_1 (m_1 a) } a_{m_1,m_2},
\end{equation}
where $\alpha$ denotes the sublattice $A,B$ or $C$; $N_1$ is the number of repeated superunit cells making up the cylinder; and integer $m_2$ indicates the triangular plaquette along the superunit cell, as shown in Fig. \ref{fig:k_BS}. The problem has become to diagonalize the one-dimensional Hamiltonian describing a superunit cell:
\begin{eqnarray}
\label{eq:k_quasi1D}
& & \mathscr{H}_{k_1} = |t|\sum_{m_2} 2 e^{i\phi} a^\dagger_{B m_2}a_{A m_2} \cos\left(k_1 \frac{a}{2}\right) \nonumber \\
& & + e^{i\phi + i k_1\frac{a}{4}} a^\dagger_{Cm_2} a_{Bm_2} + e^{i\phi - i k_1 \frac{a}{4}}a^\dagger_{Cm_2}a_{B,m_2+1} \nonumber \\
& & + e^{i\phi + i k_1\frac{a}{4}} a^\dagger_{Am_2} a_{Cm_2} + e^{i\phi - i k_1 \frac{a}{4}}a^\dagger_{Am_2}a_{C,m_2-1} + h.c. \nonumber \\
\;
\end{eqnarray}
This form reduces to the one in Eq. (\ref{eq:k_Hk}) when Fourier transformed around the remaining direction $k_2$. To obtain the edge states, let us take the general form of an eigenstate
\begin{equation}
|\Psi(k_1)\rangle \equiv \sum_{\alpha=A,B,C}\sum_{m_2} \psi_{\alpha m_2}(k_1) \ket{\alpha m_2},
\end{equation} 
where $\ket{\alpha m_2}$ is a ket localized on sublattice $\alpha$ of the $m_2^{th}$ triangular plaquette along the super-unit cell. The Schr\"odinger equation\begin{equation} 
\mathscr{H}_{k_1}|\Psi(k_1)\rangle =E_\Psi(k_1) |\Psi(k_1)\rangle
\end{equation}
yields the following linear system for the components of the wavefunction $\psi_{\alpha m_2} = \psi_{\alpha m_2}(k_1)$ and the energy dispersion $U=U(k_1)\equiv - \left(E(k_1) -\hbar \omega\right)/|t|$,
\begin{widetext}
  \begin{eqnarray}
    \label{eq:syst}
    e^{i k_1 \frac{a}{4} + i\phi} \psi_{C,m_2-1}  + 2\cos\left( k_1 \frac{a}{2}\right) e^{-i\phi} \psi_{Am_2} + e^{-i k_1 \frac{a}{4} + i\phi} \psi_{Cm_2} &=& -U \psi_{Bm_2}  \\
    e^{-i k_1 \frac{a}{4} - i\phi} \psi_{C,m_2-1} + 2 \cos\left(k_1 \frac{a}{2}\right) e^{i\phi} \psi_{Bm_2} + e^{i k_1 \frac{a}{4}-i\phi} \psi_{Cm_2} &=& -U \psi_{Am_2}  \\
    e^{ i k_1 \frac{a}{4}-i \phi} \psi_{Bm_2} + e^{-i k_1 \frac{a}{4} + i \phi} \psi_{Am_2} + e^{ -i k_x \frac{a}{4} - i \phi }\psi_{B,m_2+1}  + e^{ i k_1 \frac{a}{4} + i \phi } \psi_{A,m_2+1}  &=& -U \psi_{C,m_2} \\
    e^{-i k_1 \frac{a}{4} - i\phi} \psi_{B0}  + e^{i k_1 \frac{a}{4} + i\phi} \psi_{A0} &=& -U \psi_{C,-1}  \\
    e^{i k_1 \frac{a}{4} + i \phi} \psi_{C,m_{2\text{max}}} + 2 \cos\left(k_1 \frac{a}{2}\right) e^{ -i\phi } \psi_{Am_{2\text{max}}} &=& -U \psi_{Bm_{2\text{max}}} \\
    e^{-i k_1 \frac{a}{4} - i\phi} \psi_{Cm_{2\text{max}}} + 2 \cos\left(k_1 \frac{a}{2}\right) e^{i\phi} \psi_{Bm_{2\text{max}}} &=& -U \psi_{Am_{2\text{max}}},
  \end{eqnarray}
\end{widetext}
\noindent the first three of which hold for $m_2$ spanning the superunit cell ``bulk'' between $0$ and $m_{2,\text{max}}$, and the other three are equations at the boundary.  There exist solutions which are exponentially suppressed with distance from the boundary of the sample, which we put in the form ($\alpha$ denotes sublattice)
\begin{equation}
\psi_{\alpha m_2} = \lambda^{m_2} \psi_{\alpha 0}.
\end{equation}
The condition that the two edges of the cylinder are line-shaped amounts to requiring 
\begin{equation}
\psi_{C,-1}=\psi_{C,m_{2\text{max}}} = 0.
\end{equation}
A pair of chiral edge modes localized at the boundaries of the sample arises: 
\begin{eqnarray}
\label{eq:k_edgeStateEnFn}
E_+ = 2|t| \cos{\left(\frac{k_1 a}{2} \right)} \;\;&\text{and}&\;\;\lambda_+ = - \frac{\cos{\left( \frac{k_1}{4} - \frac{3\phi}{2}\right)}}{\cos{\left(\frac{k_1}{4} +\frac{3\phi}{2}\right)}}, \nonumber \\
E_- = - 2|t| \cos{\left( \frac{k_1 a}{2} \right) } \;\;&\text{and}&\;\; \lambda_- = \frac{\sin{\left( \frac{k_1}{4} - \frac{3\phi}{2}\right)}}{\sin{\left(\frac{k_1}{4} +\frac{3\phi}{2}\right)}}.
\end{eqnarray}
\noindent If $|\lambda_\pm|<(>)1$, then the state is confined to the top (bottom) boundary. The states identified to be at the top (bottom) travel in the $+(-)\hat{\mathbf{\Delta}}_1$ direction, so ``chirally''. The edge modes along with the bulk states for a flat band system with $\phi = \frac{\pi}{6}$ and for a dispersive middle band system at $\phi = \frac{\pi}{4}$ are plotted in the bottom panels of Fig. \ref{fig:k_BS}. The simple boundary condition chosen gives an energy dispersion similar to that of a one-dimensional tight-binding chain of lattice constant $\frac{a}{2}$. The factors $\lambda_\pm$ describing the spatial suppression of the wavefunction into the sample bulk can become singular or 0, in which case the normalized wavefunction has 0 weight inside of the bulk and is pinned to one of the two edges (see Figs. \ref{fig:k_BS} and \ref{fig:k_lambdas}). For other boundary conditions (see Fig. \ref{fig:k_bc} for the equivalent ``armchair'' condition), one can arrange that the edge mode and the middle bulk band overlap in energy, but not in momentum, or in both energy and momentum. States overlapping in energy can scatter into each other: in the ``line'' boundary conditions, a spatially localized, $\delta$-function, impurity can scatter the edge mode into the bulk, as we show in Sec. \ref{k_current}.

For the case of a Kagome ladder, the counter-propagating edge modes exist as long as there are two distinct edges, of either the ``armchair'' or the ``line'' type. Therefore, any ladder will need to have at least one row of hexagonal plaquettes. The configuration with the least sites that sustains edge modes is shown in the bottom left of Fig. \ref{fig:k_bc}. This specific ladder with two ``line'' boundaries has the property that the flux per unit cell is 0. 
 
An intriguing property related to the bulk-edge correspondence is the phenomenon of polarization in a topologically non-trivial system. We develop on this in the context of a photonic system in Sec. \ref{k_polarization}.

\section{Observables}
\label{k_obs}
In this section, we elaborate on observable quantities in the cQED based photon lattice exhibiting two different Hall phases for light \cite{Kagpho}. In Sec. \ref{k_polarization} we discuss the polarization of a topologically nontrivial bulk in the context of photonic systems and draw connections to the local density of states in Sec. \ref{k_ldos}. Sec. \ref{k_current} is dedicated to the lattice current density.  Let us emphasize that the local density of states and lattice current
densities are in principle accessible in square geometries.

\subsection{Polarization}
\label{k_polarization}
The phenomenon of polarization is a signature of the fact that it is impossible to choose a smooth gauge for the Bloch wavefunctions across the Brillouin zone in a topologically non-trivial system. Polarization has been introduced in the context of electronic Chern insulators in Ref. \cite{Vanderbilt}, and it is related to the Hall conductivity. In a photonic or cold atom system, polarization may be observed in the dynamics of wavepackets, as we are about to describe.
 
Let us consider the one-dimensional system described by the Hamiltonian $\mathscr{H}_{k_1}$ of Eq. (\ref{eq:k_quasi1D}). This time we close the cylinder around both directions $\hat{\mathbf{\Delta}}_1$ and $\hat{\mathbf{\Delta}}_2$ into a torus, such that $\mathscr{H}_{k_1}$ now describes a periodic one-dimensional lattice along the $\hat{\mathbf{\Delta}}_2$ direction, in terms of the periodic parameter $k_1$. It is possible to find for this one-dimensional periodic system a set of Bloch states $|n \mathbf{k}\rangle=|n\;k_2;k_1\rangle$ with a smooth choice of gauge everywhere as a function of $k_2$ and for each value of $k_1$. This is equivalent to the possibility to construct Wannier states which are exponentially localized along the $\hat{\mathbf{\Delta}}_2$ direction, which is a classic result of Kohn \cite{Kohn}. For now let us focus on a single band $|n \mathbf{k}\rangle \equiv | k_1 k_2 \rangle$. The maximally localized Wannier states take the following form \cite{VanderbiltKingSmith}:
\begin{eqnarray}
\label{eq:k_WannierState}
&&|W(k_1,R_2)\rangle =  \\
&&\frac{1}{\sqrt{N_2}}\sum_{k_2} e^{-i \int_0^{k_2}d\kappa_2 \mathscr{R}_2 (k_1,\kappa_2)} e^{-i k_2 (R_2-l(k_1)/2\pi)} |k_1, k_2\rangle, \nonumber
\end{eqnarray}
where $N_2$ is the number of triangular plaquettes in the quasi one-dimensional system; $R_2$ is the real coordinate of the center of the Wannier function along the $\hat{\mathbf{\Delta}}_2$ direction, which is made precise below; $\mathscr{R}_{1,2}(k_1,k_2)$ are the components along the directions $\hat{\mathbf{\Delta}}_1$,$\hat{\mathbf{\Delta}}_2$ of the Berry gauge field introduced in Eq. (\ref{eq:k_BerryGauge}), and
\begin{equation}
l(k_1) \equiv a \int_0^{\frac{4\pi}{\sqrt{3}a}} d\kappa_2 \mathscr{R}_2(k_1,\kappa_2).
\end{equation}
Since the Bloch functions can and have been chosen periodic and smooth in the $k_2$ direction, both of the components of the Berry gauge field $\mathscr{R}_{1,2}(k_1,k_2)$ are periodic functions of $k_2$.

The coordinate of the center of the Wannier functions drifts depending on the parameter $k_1$. The expectation value of the position operator $r_2$ is
\begin{equation}
\langle W(k_1,R_2) | r_2 | W(k_1,R_2) \rangle = R_2 - \frac{l(k_1)}{(4\pi/\sqrt{3})}.
\end{equation}
The net displacement of the center of the Wannier state when the parameter $k_1$ winds around its period equals the Chern number:
\begin{eqnarray}
\Delta r_2 &=& -\frac{1}{(4\pi/\sqrt{3})}\left(l\left(4\pi/\sqrt{3}a\right) - l(0) \right)  \nonumber \\
&=&  -\frac{a}{(4\pi/\sqrt{3})} \int_0^{\frac{4\pi}{\sqrt{3}a} } d\kappa_2 \left(\mathscr{R}_2\left(4\pi/\sqrt{3}a,\kappa_2\right) - \mathscr{R}_2\left(0,\kappa_2\right) \right) \nonumber \\
&=&- \frac{a}{(4\pi/\sqrt{3})} \oint_{\partial \text{BZ}} d\mathbf{k}.\mathscr{R}(k_1,k_2) = -\nu a.
\end{eqnarray}
In the last line we have used the fact that $\mathscr{R}_{1,2}(k_1,k_2)$ is periodic in $k_2$ to reexpress the drift of the center of the Wannier state as the line integral of the Berry gauge field around the boundary of the first Brillouin zone $\partial BZ$, which is exactly the Chern number $\nu$. This condition summarizes the fact that a smooth gauge cannot be chosen for the Bloch functions in a band with non-zero $\nu$. If this were true, then the Berry gauge field would be periodic in both directions, and the Chern number would vanish. We arrive at the following boundary condition for the Wannier states,
\begin{equation}
\left|W\left(k_1+4\pi/\sqrt{3}a, r_2\right)\right\rangle = |W(k_1, r_2- \nu a)\rangle.
\end{equation}
In particular, this tells us that if a wavepacket constructed out of such states is accelerated in the $\hat{\mathbf{\Delta}}_1$ direction, then its center will drift along the $\hat{\mathbf{\Delta}}_2$ direction. A wavepacket constructed within a band of $0$ Chern number would return to its initial coordinate on the $r_2$ axis upon a complete revolution of $k_1$. We shall come back to this point in Sec. \ref{k_berryphase}.

\subsection{Local density of states} 
\label{k_ldos}
\begin{figure}[t!]
\centering
\subfigure[]{
\includegraphics[width=0.45\linewidth]{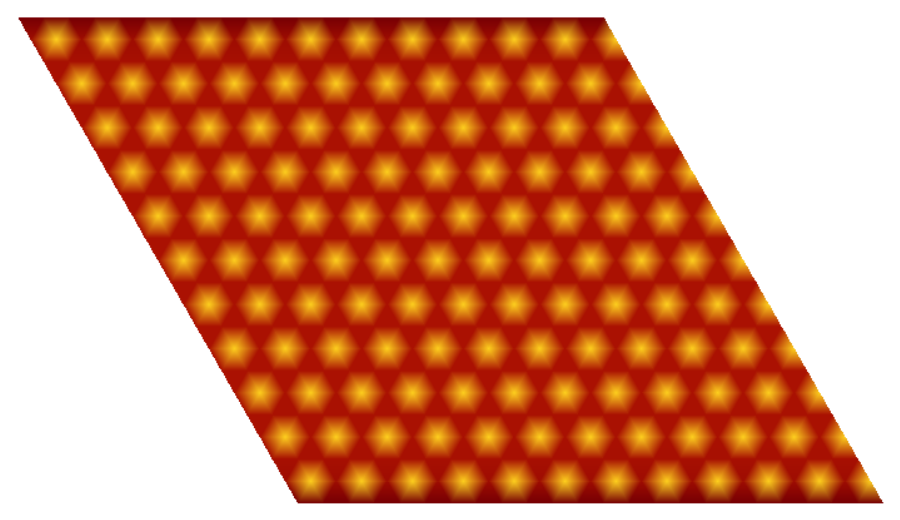}\label{fig:k_ldos_pi6clean_mid_band}}
\subfigure[]{
\includegraphics[width=0.45\linewidth]{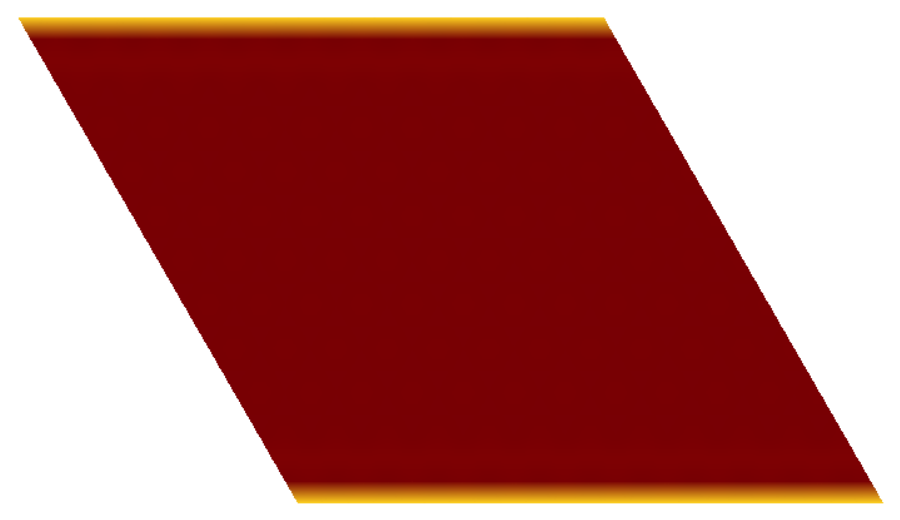}\label{fig:k_ldos_pi6clean_edge_mode}}

\vspace{0.05cm}

\subfigure[]{
\includegraphics[width=0.45\linewidth]{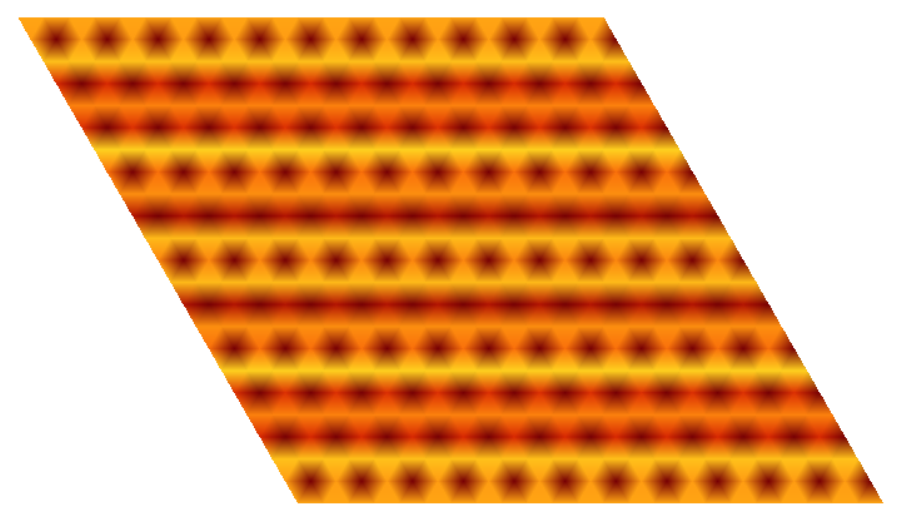}\label{fig:k_ldos_pi4clean_mid_band_0.400}}
\subfigure[]{
\includegraphics[width=0.45\linewidth]{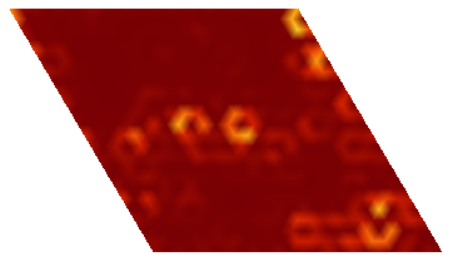}
\label{fig:k_ldos_pi6_Wphi0.94_mid_band_0.541}}
\caption{\label{fig:k_ldos_plots} (Color online) Local density of states: states are localized in the hexagonal plaquettes when the energy overlaps with the flat band at $\phi=\frac{\pi}{6}$ (\ref{fig:k_ldos_pi6clean_mid_band}); for the same system, intra-gap states are confined at the edges of the sample (\ref{fig:k_ldos_pi6clean_edge_mode}); dispersive middle band for $\phi=\frac{\pi}{4}$ contains extended states (\ref{fig:k_ldos_pi4clean_mid_band_0.400}); for comparison, local density of states for the disordered $\phi=\frac{\pi}{6}$ system is shown in \ref{fig:k_ldos_pi6_Wphi0.94_mid_band_0.541} for an energy $E$ within the proximity of the middle band.}
\end{figure}

The localization of chiral modes at the edges of the sample, the existence of a flat band, and the extended versus localized character of bulk states can be tested by measuring the local density of states. Because phase information is absent from the density, it is not possible to measure the polarization of Sec. \ref{k_polarization}. Given a complete set of states $|\Psi\rangle$, each of which is an eigenstate of energy $E_\Psi$ of the tight binding Hamiltonian of Eq. (\ref{eq:k_H}), 
\begin{equation}
\mathscr{H} |\Psi\rangle \equiv E_\Psi |\Psi\rangle,
\end{equation} 
one can define the local density of states in terms of the energy $E$ and the two dimensional real coordinate on the lattice $\mathbf{r}$
\begin{equation}
\rho( E, \mathbf{r} ) = \sum_\Psi \delta(E-E_\Psi) |\langle \mathbf{r} |  \Psi  \rangle|^2.
\end{equation}
This quantity is by definition normalized to the number of sites in the system,
\begin{equation}
\int dE \sum_{\mathbf{m}} \rho(E, \mathbf{r}_\mathbf{m}) = 3N,
\end{equation}
\noindent where $\mathbf{m} =(m_1,m_2)$ indexes the sites, $\mathbf{r}_\mathbf{m}=m_1 \mathbf{\Delta}_1+m_2 \mathbf{\Delta}_2$, and $N$ denotes the number of unit cells and there are $3$ sites per unit cell. We plot the local density of states in the same cylinder geometry with ``line'' edges of Sec. \ref{k_edgemodes}. We have approximated the $\delta$-function by a Lorentzian 
\begin{equation}
\lim_{\Delta_E\rightarrow 0} \frac{1}{\pi} \frac{\Delta_E}{(E - E_\Psi) + \Delta_E^2} = \delta( E - E_\Psi),
\end{equation}
where the width is set to $\Delta_E / |t| = 0.01$ and a system of $23\times23$ sites has been numerically diagonalized. In the flat band model with $\phi = \frac{\pi}{6}$, as shown explicitly in \cite{Kagpho}, there exists a highly degenerate band of zero-energy states localized in the hexagonal plaquettes due to destructive interference effects. When the flux $\phi$ is detuned from $\frac{\pi}{6}$, the middle band becomes dispersive and the states leak out of the hexagonal plaquettes and become extended (for $\phi=\frac{\pi}{4}$, see Fig. \ref{fig:k_ldos_pi4clean_mid_band_0.400}). Finally, if disorder is introduced into the flat band system at $\phi=\frac{\pi}{6}$, the states inside the middle band become delocalized due to disorder (Fig. \ref{fig:k_ldos_pi6_Wphi0.94_mid_band_0.541}). Intra-gap edge modes have non-zero spectral weight only close to the edges of the sample. 

\subsection{Lattice currents}
\label{k_current} 
A measurement of the local density of states $\rho(E,\mathbf{r})$ cannot  detect the chirality of edge modes. The simplest possible quantity that can show us an observable effect of chirality is the lattice current density. This operator is defined between two lattice sites indexed by two pairs of integers $\mathbf{m}$ and $\mathbf{n}$, and situated at $\mathbf{r}_{\mathbf{m}}$ and $\mathbf{r}_{\mathbf{n}}$. It measures the number of particles that flow from one site to another per unit time and takes the standard form \cite{BookWen}
\begin{equation}
j_{\mathbf{m}\mathbf{n}} = - i c^\dagger_\mathbf{m} (t_{\mathbf{m}\mathbf{n}} + t^*_{\mathbf{n}\mathbf{m}})c_\mathbf{n} + i c^\dagger_\mathbf{n}(t_{\mathbf{m}\mathbf{n}}^* + t_{\mathbf{n}\mathbf{m}})c_\mathbf{m}.
\end{equation}
Using the notation of Sec. \ref{k_ldos}, the observable quantity of interest is the current expectation value for an eigenstate $|\Psi\rangle$ of the tight-binding Hamiltonian $\mathscr{H}$ is $\langle \Psi | j_{\mathbf{m}\mathbf{n}} |\Psi \rangle$.

From the study of edge mode dispersion relations in Sec. \ref{k_edgemodes}, we have concluded that the introduction of a small impurity at the boundary can mix all momenta and the edge state can scatter into the bulk. We test this by introducing a $\delta$-function impurity at one of the sites on the boundary of the system.  The expectation value $\bra{\Psi} j_{\mathbf{m}\mathbf{n}} \ket{\Psi}$ is the familiar current density associated with the wavefunction $\ket{\Psi}$. 

\begin{figure}[t!]
\begin{center}
\includegraphics[width=0.49\linewidth]{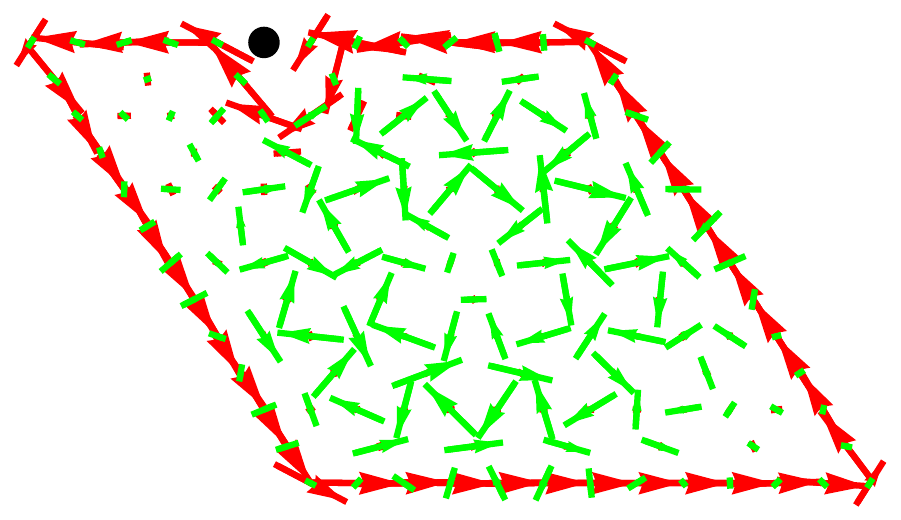}
\includegraphics[width=0.49\linewidth]{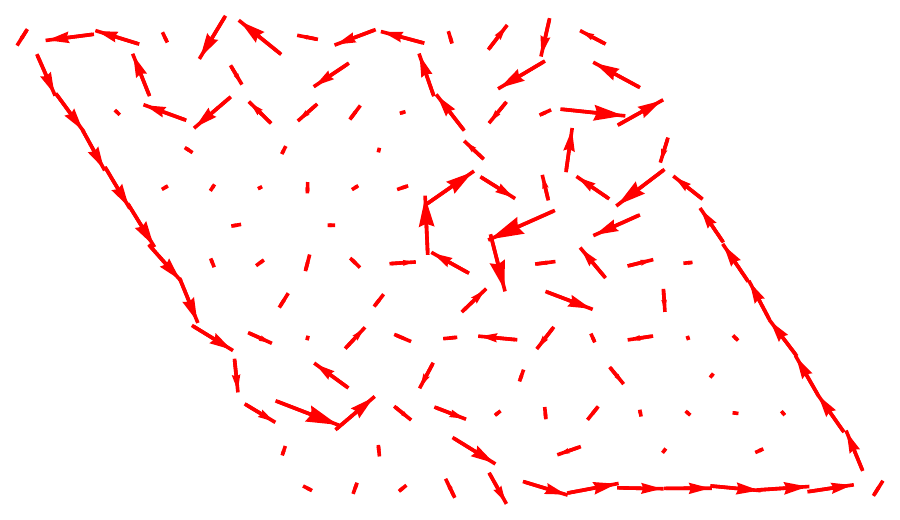}
\caption{\label{fig:k_Currents} (Color online) Plots of the expectation value of the lattice current operator in two implementations of the $\phi = \frac{\pi}{4}$ system. Left panel: the system has a dispersive middle band overlapping in energy with the edge mode (See Fig. \ref{fig:k_bc}). In the presence of a $\delta$-function impurity the edge state current (red) will deviate around the impurity, maintaining its chirality, if the energy of the edge state lies in the gap and does not overlap with the middle band. This corresponds to the anomalous quantum Hall phase. The current for a state whose energy is within the overlap region with the middle band is shown in green. Right panel: in the presence of a dispersive middle band (with $\phi = \frac{\pi}{4}$) and disorder, the edge state can leak into the bulk with a finite probability.}
\end{center}
\end{figure}

We consider a system containing one impurity localized in real space at a single site. This impurity couples all pairs of $\mathbf{k}$-points, implying that whenever the edge mode and the bulk state are at the same energy, but separated in momentum, the edge mode will have a finite lifetime towards scattering into a bulk state. Results for the current operator in the Kagome system are plotted in Fig. \ref{fig:k_Currents}. In the left panel, we have plotted the current of an edge state whose energy is located in the gap just above and not overlapping with the middle band at $\phi = \frac{\pi}{4}$ (see Fig. \ref{fig:k_BS}) - the current (in red) will go around the $\delta$-function impurity without leaking into the bulk. This corresponds to the anomalous quantum Hall phase and is reminiscent of the situation at $\phi=\frac{\pi}{6}$. As soon as the edge mode and the  bulk band overlap in energy, current at the edge has a probability to scatter into the bulk. A state at this energy will essentially live inside of the bulk (current depicted in green). This corresponds to the anomalous Hall phase. If instead of an impurity at the edge we had strong phase disorder $W_\phi = \frac{\pi}{3}$ for a system with flux $\phi = \frac{\pi}{4}$, there would again be a finite probability to scatter into the bulk, as shown in the right panelof Fig. \ref{fig:k_Currents}.

\section{Berry's phase and Anomalous Hall effect of neutral particles}
\label{k_berryphaseobs}
When a system in an anomalous Hall phase is placed in an external gauge field, it is possible to measure Berry's phases around closed loops in the Brillouin zone and band Chern numbers. This section is dedicated to such observables. The treatment is applicable whenever the particles are neutral, and therefore suitable for cold atom systems as well. In Sec. \ref{k_synthetic} we explain a scheme to realize synthetic gauge fields on the Kagome lattice that involves a time-modulated potential gradient in the cavity array, as suggested in Ref. \onlinecite{Kolovsky}. In Sec. \ref{k_berryphase}, we present a way to measure the Berry phases of wavepackets of photons at a given energy based on interference. We also present a manner in which the Chern number can be directly extracted from level-counting in the spectrum of the system in a uniform synthetic magnetic field in Sec. \ref{k_directchern}.

\subsection{Realization of a synthetic gauge field}
\label{k_synthetic}
We discuss here how to simulate the effect of a magnetic field in a system of neutral particles. Such artificial magnetic fields, including frustrated or staggered configurations, have been actively studied in recent years. In a cavity, photons coupled to artificial atoms, such as in the Rabi model, are subject to an effective gauge potential induced by the dipolar coupling \cite{LarsonLevin}. On a lattice,  gauge fields can in principle be simulated by creating effective Peierls phases through time-periodic driving as proposed in Refs. \cite{Kolovsky,HaukeEtAl,StruckEtAl} for cold atoms. In what follows, we shall follow the line of thought of Ref. \cite{Kolovsky}. The method relies on adjusting  the potential terms individually for each site in the trapping optical lattice of the cold atom system.  In cQED systems, this is equivalent to adjusting the frequency of each resonator individually and time-dependently, which has become experimentally possible \cite{Greentree2,Johansson, Sandberg,DelsingPRL}. Ref. \cite{Greentree2} has already proposed a means of realizing a synthetic gauge field in a square lattice array of coupled photonic cavities.

Generally, a static gradient of the frequency $\omega_\mathbf{m}$ of the resonator located at site $\mathbf{m}$ yields an artificial electric field. Additionally, it is possible to have time-varying frequencies $\omega_{\mathbf{m},\tau}$, which produce complex hopping amplitudes and so mimic the effect of a gauge field. Consider that that there is a driving frequency at site $\mathbf{m}=(m_1,m_2)$ with the following dependence on the time $\tau$:
\begin{equation}
\label{eq:k_Htau}
\mathscr{H}_\tau = \mathscr{H} + \sum_\mathbf{m} \left(\hbar \omega+ \hbar\omega_0 \cos \left( \Omega \tau + m_2 \theta  \right) m_1 \right) a^\dagger_\mathbf{m} a_\mathbf{m},
\end{equation}
\noindent where the time-dependent Hamiltonian is the original tight-binding Hamiltonian of Eq. (\ref{eq:k_H}), to which we add a time-dependent perturbation. The perturbation amounts to a tilt along the $\hat{\mathbf{\Delta}}_1$ direction, modulated in time. $\Omega$ is the driving frequency, and for the following derivation we assume that the driving frequency is resonant with the on-site energy $\hbar \omega$ in the sense that $\Omega = \omega$. The details of this choice and the calculation are presented in Appendix \ref{ap:synthetic}. Here we summarize the results.

In the rotating wave approximation, the time dependence of Eq. (\ref{eq:k_Htau}) has the following effects: photons acquire an additional phase $m_2 \theta$ along each bond in the $\hat{\mathbf{\Delta}}_1$ direction; they acquire no additional phase along the bonds in the $\hat{\mathbf{\Delta}}_2$ direction. The following equations summarize the changes incured by the complex hopping amplitudes:
\begin{eqnarray}
&& \text{bond} \parallel \hat{\mathbf{\Delta}}_1:  |t| e^{-i \phi} \rightarrow |t| e^{-i \phi}\; \cdot\; \mathcal{J}_{-1}\left( \frac{\omega_0}{\Omega} \right) e^{-i{m_2 \theta}}, \nonumber \\
&& \text{bond} \parallel \hat{\mathbf{\Delta}}_2:  |t| e^{-i \phi} \rightarrow |t| e^{-i \phi} .
\end{eqnarray}
\noindent The function $\mathcal{J}_{-1}(z)$ is a Bessel function of the first kind. There will be a total phase $f=2\theta$ around the parallelogram plaquettes of area $|\mathbf{\Delta}_1 \times \mathbf{\Delta}_2|$ (see Fig. \ref{fig:k_BS}). The artificial gauge field extracted from Eq. (\ref{eq:k_Htau}) gives rise to a uniform magnetic flux across the lattice, although the field itself is not uniform at the level of one unit cell due to the presence of the oblique bonds; nevertheless, this non-uniformity can be removed by a gauge transformation (see Appendix \ref{ap:synthetic}). This construction is equivalent to the Peierls substitution: a particle in a gauge field $\mathscr{A}_s(\mathbf{r})$ acquires a phase $\theta_{\mathbf{m}\mathbf{n}}=\frac{e_s}{\hbar}\int_\mathbf{r_\mathbf{m}}^{\mathbf{r}_\mathbf{n}} d\mathbf{r}.\mathscr{A}_s(\mathbf{r})$ between two sites located at $\mathbf{r}_\mathbf{m}$ and $\mathbf{r}_\mathbf{n}$ (the choice of path does not matter if the vector potential is that of an infinitely thin solenoid at the center of the plaquette). We shall make the convention that the photon coupling to such artificial gauge fields has an effective synthetic charge $e_s$, and all fields to be used below will have a subscript ``\textit{s}'' to highlight the distinction from actual electromagnetic fields. 

The experimental realization of an artificial field would allow us to measure Berry's phases around closed loops in the Brillouin zone or directly access the Chern number, as explained in Sec. \ref{k_berryphase} and \ref{k_directchern}, respectively. The potential to create a magnetic field can make a measurement of Landau levels of the Kagome lattice accessible, see Sec. \ref{k_directchern}.

\subsection{Berry's phases from semiclassical dynamics}
\label{k_berryphase}
In this section we describe a method to measure Berry's phases around closed constant energy contours in the Brillouin zone. For this we draw upon the semiclassical dynamics of a wavepacket within a given band of a Bloch Hamiltonian. 

In Sec. \ref{k_polarization} we have noted that, due to the polarization phenomenon, wavepackets accelerated in the $\hat{\mathbf{\Delta}}_1$ direction drift in the $\hat{\mathbf{\Delta}}_2$ direction. The motion of wavepackets subject to a uniform force is complicated due to Bloch oscillations. Proposals exist to access Berry's phases in a cold atom system relying on a measurement of the group velocity of a wavepacket under an external scalar potential exhibiting Bloch oscillations \cite{PriceCooper}. Here we present an alternative method to map Berry's phases of wavepackets, based solely on interference, and without the need to measure the group velocity of wavepackets. 

The semiclassical dynamics of a wavepacket centered at momentum and coordinate $\mathbf{k}_c, \mathbf{r}_c$ and in a Bloch band is given by \cite{Marder,ChangNiu,Polkovnikov}
\begin{eqnarray}
\label{eq:k_SemiCls}
\dot{\mathbf{r}}_c &=& \frac{1}{\hbar} \frac{\partial E_{\mathbf{k}_c}}{\partial \mathbf{k}_c} - \dot{\mathbf{k}}_c \times \mathscr{F}_{\mathbf{k}_c} \nonumber \\
\hbar \dot{\mathbf{k}}_c &=& - e_s \mathscr{E}_s(\mathbf{r}_c) - e_s \dot{\mathbf{r}}_c \times \mathscr{B}_s(\mathbf{r}_c) ,
\end{eqnarray}
\noindent where $\mathscr{E}_s,\mathscr{B}_s$ are synthetic classical fields, and $e_s$ is a synthetic effective charge, which is left as an explicit quantity in the equations in order to highlight the analogy with electronic systems. $E(\mathbf{k}_c)$ is the Bloch band energy, and  
\begin{equation}
\label{eq:k_BerryCurvature}
\mathscr{F}_{\mathbf{k}} \equiv \partial_{\mathbf{k}} \times \mathscr{R}(\mathbf{k})
\end{equation}
\noindent is the Berry curvature \cite{Berry} associated with the Berry gauge field introduced in Eq. (\ref{eq:k_BerryGauge}); we have dropped the band index $n$, as we neglect transitions to other Bloch bands; $\mathscr{F}$ only has a $\hat{\mathbf{z}}$ component. The semiclassical Eqs. (\ref{eq:k_SemiCls}) are dual in the following sense: the second equation represents the electromagnetic force; in the first equation, the band energy plays the role of a scalar potential in momentum space, while the second term is nonvanishing only when the wavepacket is accelerated, and the Berry curvature acts like a magnetic field in momentum space. The group velocity $\frac{\partial E}{\partial \mathbf{k}_c}$ is corrected by an ``anomalous'' contribution \cite{KarplusLuttinger} which is now known to be inherently connected to geometric phases of wavefunctions \cite{ChangNiu,JungwirthEA2002}. 

\begin{figure}[t!]
\begin{center}
\includegraphics[width=0.64\linewidth]{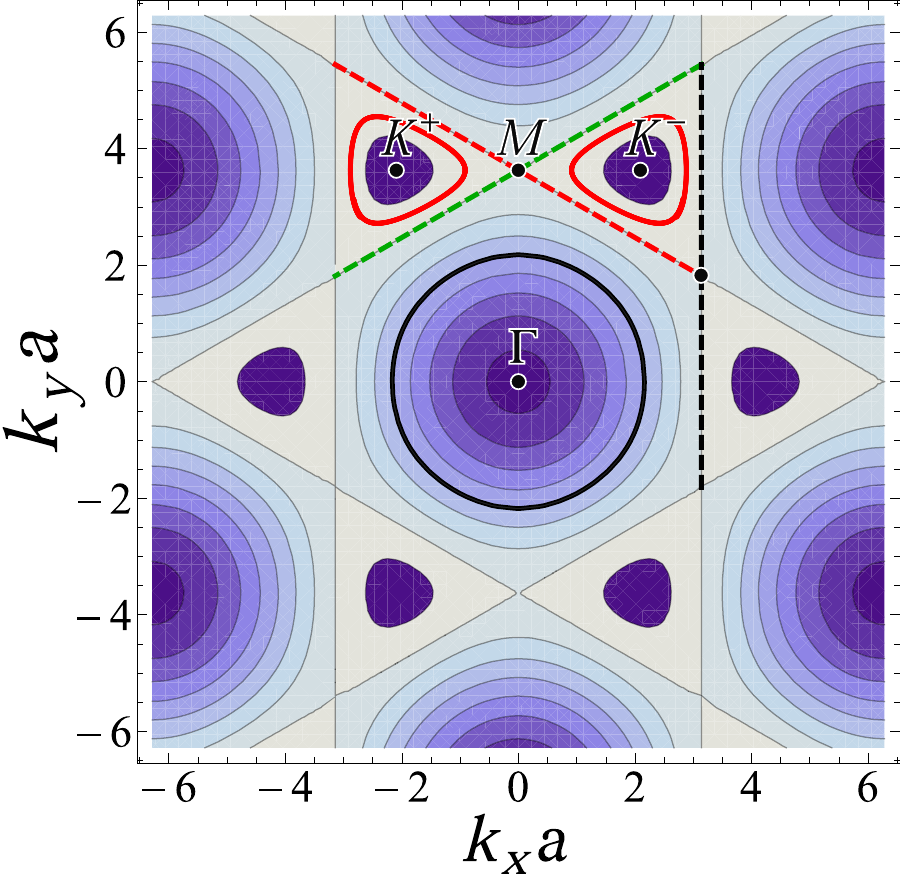}
\caption{(Color online) Constant energy trajectories in momentum space, overlayed on the constant energy contours of the lowest band at $\phi = \frac{\pi}{6}$. Trajectories can be closed within the first Brillouin zone resulting in closed real space trajectories or, if the initial momentum lies on the separatrix lines (see, for example, the dashed lines), then the trajectory in momentum space can only close in the extended Brillouin zone, and in real space the particle traces a straight line. \label{fig:k_SclPaths}}
\end{center}
\end{figure}

The geometric phase of a wavepacket can be measured in a system subject to a uniform synthetic magnetic field perpendicular to the plane, $\mathscr{B}_s = |\mathscr{B}_s| \hat{\mathbf{z}}$, but no electric field. The semiclassical equations of motion for the wavepacket become
\begin{eqnarray}
\dot{\mathbf{r}}_c &=& \frac{\hbar}{ e_s \mathscr{B}_s } \dot{\mathbf{k}}_c \times \hat{\mathbf{z}} \label{eq_rcdot} \label{eq:k_rcEq} \\
\hbar \dot{\mathbf{k}}_c &=& - \frac{e_s}{\hbar} Z_{\mathscr{B}_s} (\mathbf{k}_c) \frac{\partial E}{\partial \mathbf{k}_c} \times \mathscr{B}_s \label{eq_kcdot} \label{eq:k_kcEq},
\end{eqnarray}
\noindent and $Z_{\mathscr{B}_s}$ is a correction from the Berry curvature:
\begin{equation}
\label{eq:k_BerryCorr}
Z_{\mathscr{B}_s} (\mathbf{k}_c) = \frac{1}{1 + \frac{e_s}{\hbar} \mathscr{F}_{\mathbf{k}_c} \cdot \mathscr{B}_s}.
\end{equation}
\noindent Eq. (\ref{eq:k_kcEq}) implies that the force $\dot{\mathbf{k}}_c$ is always perpendicular to the group velocity $\frac{\partial E}{\partial \mathbf{k}_c}$. Consequently, the energy of the wavepacket is a constant of motion, 
\begin{equation}
\delta E(\mathbf{k}_c) = \frac{\partial E}{\partial \mathbf{k}_c} .\delta \mathbf{k}_c = 0.
\end{equation}
\noindent The trajectory in momentum space follows a contour of constant energy in the Brillouin zone. The trajectory in real-space described by Eq. (\ref{eq:k_rcEq}), on the other hand, is related to the trajectory in momentum space by a $\frac{\pi}{2}$ rotation and a rescaling. A wavepacket evolving on a constant energy contour will return to the original point in $\mathbf{k}$-space. If the orbit is closed in the first Brillouin zone (continuous lines in Fig. \ref{fig:k_SclPaths}), then the trajectory in real space will be a closed curve as well. Alternatively, the orbit may be periodic in the extended Brillouin zone and follow a separatrix (dashed lines in Fig. \ref{fig:k_SclPaths}), in which case the trajectory in real space will not be closed. Finally, the Berry curvature correction in Eq. (\ref{eq:k_BerryCorr}) can only affect the rate at which the momentum $\mathbf{k}_c$ varies, but not the form of the trajectory. The acceleration will be small in areas of strong Berry curvature, typically close to the $K$ points.

It is possible to determine resonance conditions for those paths which close in the Brillouin zone. Let us assume that we are in the case in which the wavepacket traverses a closed curve in phase space. Along a phase space loop of constant energy $E$, denoted $C(E)$, the wavepacket will acquire a phase \cite{ChangNiu,Marder}
\begin{equation}
\label{eq:k_gammaC}
\gamma_C = \oint_C d\mathbf{k} \cdot \mathscr{R}(\mathbf{k})  + d\mathbf{r}.\left( \mathbf{k} - \frac{e_s \mathscr{A}_s }{\hbar}  \right),
\end{equation}
where the gauge field $\mathscr{A}_s$ generates the magnetic field $\mathscr{B}_s = \partial_\mathbf{r} \times \mathscr{A}_s$. This can be further simplified to
\begin{eqnarray}
\gamma_C &=& \oint_C d\mathbf{k} \cdot \mathscr{R}(\mathbf{k}) - \int_0^T dt \left(\frac{e_s \mathscr{A}_s }{\hbar} \dot{\mathbf{r}}  - \frac{e_s}{\hbar} (\dot{\mathbf{r}} \times \mathscr{B}_s) \cdot \mathbf{r} \right) \nonumber \\
&=& \oint_C d\mathbf{k} \cdot \mathscr{R}(\mathbf{k}) + \frac{e_s}{2 \hbar} \int_0^T dt \big( \left(\dot{\mathbf{r}} \times \mathscr{B}_s\right) \cdot \mathbf{r}  \big),
\end{eqnarray}
\noindent where $T$ is the period of the motion. The first term of the integrand is the Berry phase accumulated along the path, $\Gamma_C \equiv \oint_C d\mathbf{k} . \mathscr{R}(\mathbf{k})$. The second term can be written entirely in momentum space,
\begin{eqnarray}
&& \frac{e_s}{2 \hbar} \int_0^T dt \big( \left(\dot{\mathbf{r}} \times \mathscr{B}_s \right) \cdot \mathbf{r}  \big)  \nonumber \\
&=& \frac{e_s}{2\hbar} \int_0 ^T dt  \left( \frac{\hbar}{e_s \mathscr{B}_s} \dot{\mathbf{k}} \times \hat{\mathbf{z}} \times \mathscr{B}_s \right) \left( \frac{\hbar}{e_s \mathscr{B}_s} \mathbf{k} \times \hat{\mathbf{z}} \right) \nonumber \\
&=& \frac{\hbar}{2 e_s \mathscr{B}_s} \oint_{C} d\mathbf{k} \cdot \left(\mathbf{k} \times \hat{\mathbf{z}}\right) \nonumber \\
&=& \frac{\hbar}{e_s \mathscr{B}_s} \mathscr{S}_C,
\end{eqnarray}
\noindent where $d\mathbf{k} \cdot \left(\mathbf{k} \times \hat{\mathbf{z}}\right) \equiv 2\; d \mathscr{S}$ is twice the surface area element for the surface enclosed by the curve $C(E)$ in momentum space. Collecting the two pieces, the full phase acquired by the wavepacket around a closed loop in momentum space is
\begin{equation}
\gamma_C = \Gamma_C + \frac{\hbar}{e_s\mathscr{B}_s} \mathscr{S}_C.
\end{equation}
The arriving wavepacket will interfere with the emitted wavepacket and the amplitude will be maximal if
\begin{equation}
\label{eq:k_GammaC}
\Gamma_C + \frac{\hbar}{e_s\mathscr{B}_s} \mathscr{S}_C = 2 \pi \left( n + \nu_M \right), \;\; n\;\text{=  integer},
\end{equation}
where the Maslov index \cite{Maslov} is taken to be $\nu_M=\frac{1}{2}$. There is an implicit energy dependence in the trajectory $C=C(E)$, therefore the phases in Eq. (\ref{eq:k_GammaC}) are energy dependent. This last equation is the Onsager relation \cite{OnsagerRel} and can be viewed as a Bohr-Sommerfeld quantization condition for the closed trajectories in phase space. An analogous situation occurs in electronic systems, in the de Haas - van Alphen effect \cite{Marder}, where peaks in the magnetization as a function of the external magnetic field are a result of such a resonant behavior. 

In a photon system, the Berry phases can be accessed as follows. Suppose that a wavepacket of known energy $E$ and initial momentum $\mathbf{k}_c (t = 0)$ can be produced. Upon tracing a closed path in real space, the traveling wavepacket returns and interferes with the emitted signal.  One can tune the external field $\mathscr{B}_s$ such that the amplitude at the emitter is resonant. Two consecutive resonances determine the area $\mathscr{S}_C$ of the common closed path in momentum space, see Eq. (\ref{eq:k_GammaC}). This allows for the determination of the Berry phase along the path $C(E)$ up to a multiple of $2\pi$. By keeping the energy $E$ fixed and changing initial momentum one can explore all curves at the given energy within the Brioullin zone. Two examples are given in Fig. \ref{fig:k_SclPaths}. What such a measurement would yield with and without disorder is presented in Sec. \ref{k_chern} in connection to a realization of the anomalous Hall effect in the Kagome photon system.

We have shown that an interference experiment can be realized to  measure the Berry phase of a wavepacket injected at a specific energy and initial momentum. Berry phases appear in measurements of a photonic equivalent of the anomalous Hall effect, as explained in the next section.

Moreover, the measurement can be changed to probe Landau levels. Instead of keeping the energy fixed and varying the magnetic field $\mathscr{B}_s$, let us keep the field fixed and vary the energy. Separations between resonant orbits discussed above correspond to transitions between Landau levels. The analysis of level splitting in the presence of $\mathscr{B}_s$ allows us to directly probe the Chern number of a Bloch band. This is the subject of Sec. \ref{k_directchern}.

\subsection{Direct determination of Chern number from level counting in the Hofstadter spectrum}
\label{k_directchern}

In this section we present a way to directly determine the Chern number of a Bloch band from the spectrum in a synthetic magnetic field. In a magnetic field, the three original Bloch bands will split into subbands, sometimes called magnetic Bloch bands. The resonant semiclassical trajectories of Sec. \ref{k_berryphase} correspond to the magnetic subbands of the original Bloch band under the influence of the synthetic magnetic field $\mathscr{B}_s$. The Chern number influences how a Bloch band splits into magnetic subbands \cite{ChangNiu}. There exists a maximum number of closed resonant trajectories within the Brillouin zone. This is obtained by observing that the maximal area of a single resonant closed trajectory $C$, $\mathscr{S}_C$, has to be equal to the area of the first Brillouin zone.  This dictates, then, that the number of subbands obtained by splitting any one of the Bloch bands is given by:
\begin{equation}
\label{eq:k_splitting}
D = \left[ \nu + \frac{1}{f} \right],
\end{equation}
\noindent where $\nu$ is the Chern number associated with the band, $f = \frac{\mathscr{B}_s a^2 \sqrt{3}}{2}$ is the flux through the unit cell of a uniform synthetic magnetic field $\mathscr{B}_s$, in units of $\frac{h}{e^2}$, and the square brackets indicate the integer part of the real number. The field $\mathscr{B}_s$ has to be small enough such that the resulting subbands cannot become degenerate. This is argued in more detail numerically in what follows.

\begin{figure}[t!]
\centering
\includegraphics[width=0.95 \linewidth]{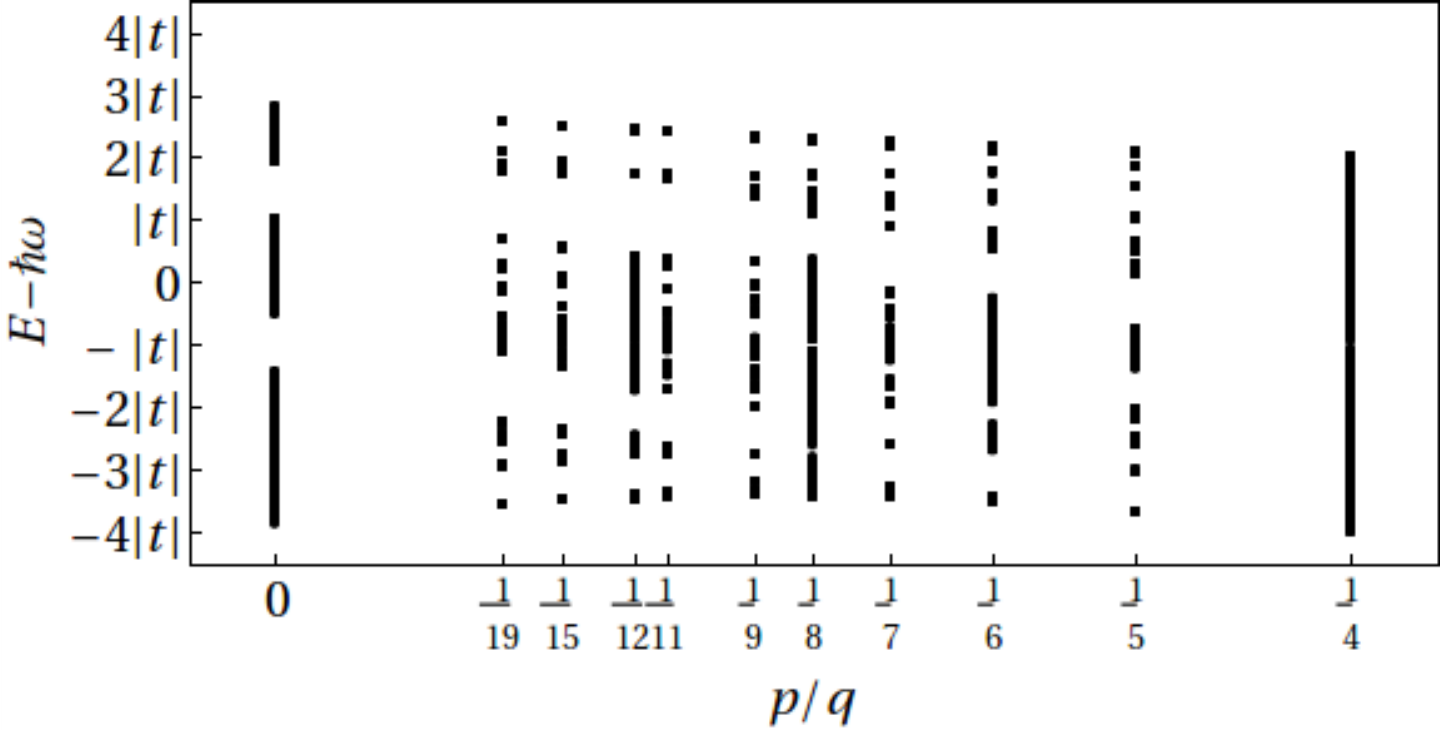}
\caption{\label{fig:k_lowfieldspectrum} Low magnetic field $\mathscr{B}_s$ spectrum of the Kagome lattice with $\phi = \frac{\pi}{4}$: at 0 external field, we recover the original bandstructure of Fig. \ref{fig:k_BS}. At flux $p/q$ there are $3q$ subbands with possible degeneracies. For example, at $p/q = 1/5$, the density of states reveals that there are 6 separated subbands in the upper band, 5 degenerate subbands in the middle band, and 4 separated subbands in the lower band in agreement with Eq. (\ref{eq:k_splitting}), and consistent with the fact that the lower and upper Bloch bands have Chern numbers -1 and 1, respectively. The bands become degenerate (no energy gaps in the spectrum) at $p/q = 1/4$. See text for details.}
\end{figure}

We are aiming to solve the Hofstadter problem of a tight-binding system in a magnetic field \cite{Hofstadter}. Let us pick the Landau gauge, $\mathscr{A}_s = ( -\mathscr{B}_s y,0)$, where the two components are cartesian. Due to the presence of a magnetic field, the Hamiltonian will couple different points in the Brillouin zone. If the flux $f$ introduced above for the uniform synthetic field is a rational number expressed as $p/q$ with $p$ and $q$ non-negative integers which are relatively prime, then, in general, the resulting Hamiltonian will couple $k_y$ to $k_y \pm \frac{4\pi f}{a\sqrt{3}}$ and to $k_y \pm \frac{8\pi f}{a\sqrt{3}}$. This momentum space coupling can be removed by remarking that translational invariance is recovered if one reverts to a $q$ times larger unit cell in real space, and a $q$ times smaller Brillouin zone, called the magnetic Brillouin zone, defined for our lattice as  $\left[0,\frac{2\pi}{q a}\right]\times \left[0,\frac{4\pi}{a\sqrt{3}}\right]$. The original $3$-band Hamiltonian becomes a $3q$-band problem defined on the reduced Brillouin zone. Folding of bands appears, and at a general level the $3q$ bands are allowed to become degenerate within the magnetic Brillouin zone. Technical details are given in Appendix \ref{ap_butterfly}. For now we focus on counting the magnetic subbands obtained from a given Bloch band, which provides a direct way of determining the Chern number of the original (zero field) bands.

Assume for simplicity that the dimensionless flux per unit cell is $f = \frac{1}{Q}$ for some positive integer $Q$. For large enough $Q$, the field is weak, and any one of the three original Bloch bands will split, according to Eq. (\ref{eq:k_splitting}), into $\left[ \nu + Q \right]$, where $\nu$ is the Chern number on each of the three original Bloch bands. This is consistent with the fact that the spectrum in a magnetic field at flux $1/Q$ should exhibit $3Q$ levels. Taking the band-structure at $\phi = \frac{\pi}{4}$ (see Fig. \ref{fig:k_BS}) as representative for our time-reversal symmetry broken phase with gapless edge modes, the formula in Eq. (\ref{eq:k_splitting}) predicts that the lower, middle and upper bands will split into $Q-1$, $Q$ and $Q+1$ subbands, respectively. The caveat to this discussion is that for special fractions $f$ one can expect that the resulting subbands become degenerate even if the original system is non-degenerate \cite{ChangNiu, Thouless}. 

One can directly access the Chern number of a specific band as follows (see Fig. \ref{fig:k_lowfieldspectrum}): set the external synthetic field to a specific rational number dimensionless flux and count the resonant trajectories corresponding to each of the three original Bloch bands (they will be typically separated by gaps of order $|t|$, as in the original system, if the field is weak). For example (see Fig. \ref{fig:k_lowfieldspectrum}), at $f=1/5$, the lower, middle, and upper Bloch bands split into $[\nu + 5] = 4,5,6$ subbands, consistent with their Chern numbers $\nu = -1, 0, 1$, respectively. There are a total of 15 subbands, but the 5 subbands corresponding to the middle band are degenerate. Importantly, at $f=1/4$, the spectrum has no energy gaps, in which case Chern numbers are exchanged between bands. We have proved numerically the requirement $f < 1/4$ to probe the Chern numbers of the original Bloch bands. Since the spectrum of the Hofstadter problem is periodic in $f$, this requirement actually translates to $f\in [0,1/4)\cup [2,9/4)\cup ...$. 

\subsection{Bounds on the strength of the synthetic magnetic field}
When the external field is too strong, or varies too quickly, tunneling to a different band becomes possible. To be safe from Landau-Zener tunneling \cite{Zener}, the magnetic field has to be small. This translates to a condition \cite{Marder} on the period of motion on a closed loop, $T$, introduced in our calculation above,
\begin{equation}
\frac{\hbar}{T} \ll E_g \sqrt\frac{E_g}{E},
\end{equation}
\noindent where $E_g$ is the size of the gap, which in our problem is of order the hopping energy, $E_g \sim |t|$, and $E$ is the constant energy along the loop $C(E)$. The period of motion can be further reexpressed by making an estimate from Eq. (\ref{eq_kcdot}),
\begin{equation}
T \sim \frac{\hbar l_C(E)}{ \frac{e_s}{\hbar} \overline{Z_{\mathscr{B}_s}} \overline{\left(\frac{\partial E}{\partial \mathbf{k}_c}\right)} .\mathscr{B}_s},
\end{equation}

\noindent where the averages are taken over the path $C$, whose length in momentum space is denoted $l_C(E)$. Collecting equations yields the following condition
\begin{equation}
\frac{\hbar l_C(E)}{e_s \overline{Z_{\mathscr{B}_s}(\mathbf{k}_c)} \overline{\left(\frac{\partial E}{\partial \mathbf{k}_c}\right)} . \mathscr{B}_s } \gg \frac{1}{|t|}\sqrt{\frac{E}{|t|}}.
\label{eq_constrB1}
\end{equation}
\noindent It turns out that this bound is independent of the lattice spacing $a$ or the absolute value of the hopping integral $|t|$. For the typical values of the Berry curvature etc., the bound of Eq. (\ref{eq_constrB1}) is satisfied by taking the flux of $\mathscr{B}_s$, $f\ll 1$, which is consistent with the results in the Sec. \ref{k_berryphase}.

\begin{figure*}[t!]
\includegraphics[width=\linewidth]{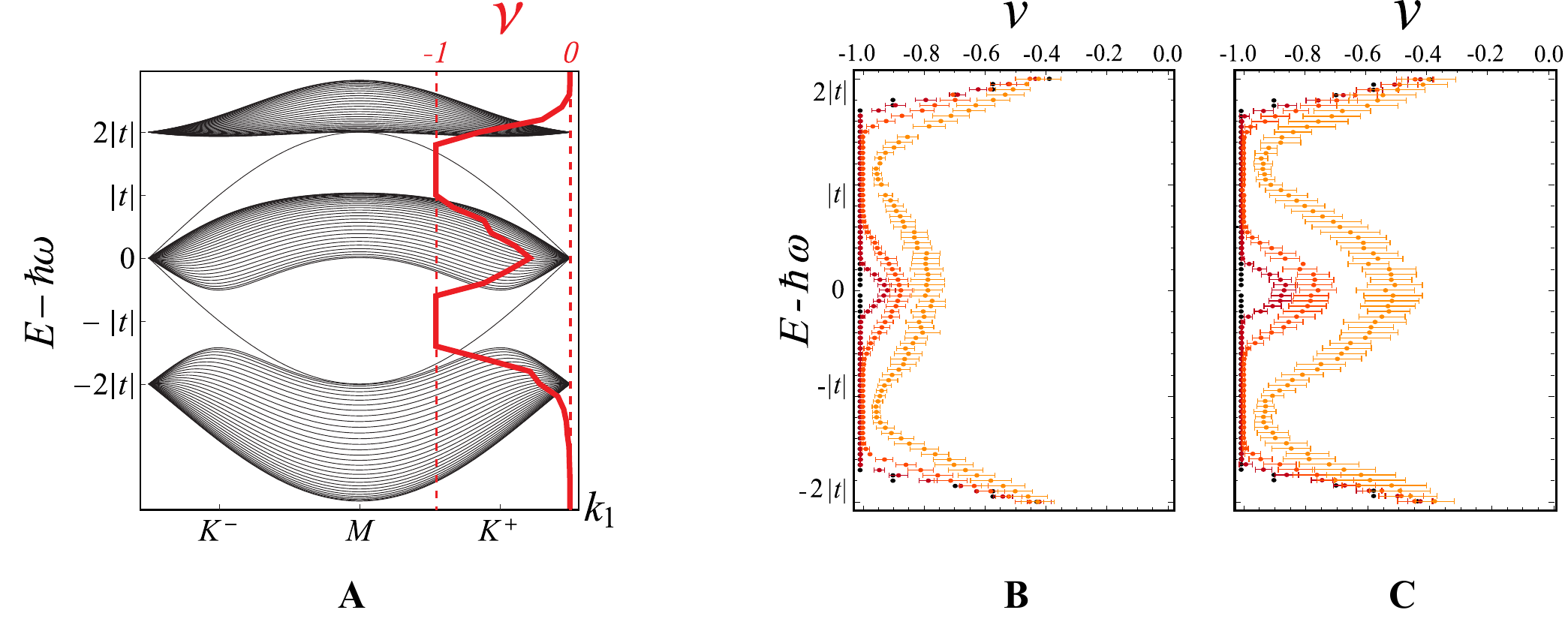}
\caption{\label{fig:k_chern} (Color online) Panel A: Dispersive middle band for $\phi = \frac{\pi}{4}$. The Chern number, computed with formula of Eq. (\ref{eq:k_nuE}) gives a non-quantized value in the overlap region which can be measured directly in an interference experiment. Panel B: Chern number calculation for the flat band system $\phi=\frac{\pi}{6}$ with increasing amplitude of phase disorder: $W_\phi = 0, \frac{\pi}{6} \cdot \frac{2}{5},  \frac{\pi}{6},  \frac{\pi}{6} \cdot \frac{9}{5}$. Panel C: flat band system with increasing amplitude of on-site disorder $W_{site} = 0, \frac{|t|}{2}, |t|, \frac{3|t|}{2}$. The disorder potential has no spatial correlation, and is drawn from a uniform distribution (see text). }
\end{figure*}

\section{Anomalous Hall Effects of Photons}
\label{k_chern}

In this section, we discuss the anomalous Hall effects of light and evaluations of Chern numbers for the clean and for the disordered case. 

\subsection{Chern number}
The experiment of Sec. \ref{k_berryphase} amounts to measuring line integrals of the Berry gauge field $\mathscr{R}(\mathbf{k})$ \cite{Berry}:
\begin{equation}
\Gamma_C(E) = \oint_C d\mathbf{k} . \mathscr{R}(\mathbf{k}).
\end{equation}

\noindent Since $C(E)$ is a constant energy curve in the Brillouin zone, the phase can be reexpressed as a sum only over the states of the Bloch band which lie below the given energy of the curve, $E$,
\begin{equation}
\Gamma_C(E) = \int_\text{BZ} d^2\mathbf{k}\; \theta\left(E-E(\mathbf{k})\right) \mathscr{F}_\mathbf{k}.  
\end{equation}
\noindent This value is, in general, not quantized: only after summing over the full Bloch band can the phase become a multiple integer of $2\pi$. In electronic systems, the non-quantized part is the intrinsic contribution to the anomalous Hall effect \cite{NagaosaEtAl,JungwirthEA2002} of a partially filled band (non-quantized Hall conductivity $\sigma_{xy}$), and can be interpreted as a Berry phase of quasiparticles at the Fermi surface \cite{Haldane2004}. In a bosonic system, we interpret this as the anomalous transport of a wavepacket whose energy overlaps with that of a bulk band. 

So far, we have dealt with a single Bloch band. In a multiple band system, we can define the following quantity as a sum over states below some energy:
\begin{equation} 
\label{eq:k_nuE}
\nu(E) = \frac{1}{2\pi} \sum_n \int_{\text{BZ}} d^2\mathbf{k} \theta(E - E_n (\mathbf{k})) \mathscr{F}^{(n)}_\mathbf{k},
\end{equation}
\noindent where now the Berry curvature $\mathscr{F}^{(n)}_\mathbf{k}$ is defined as in Eqs. (\ref{eq:k_BerryGauge}) and (\ref{eq:k_BerryCurvature}) but for the $n^{th}$ Bloch band $|n \mathbf{k}\rangle$. $\nu(E)$ is plotted in Fig. \ref{fig:k_chern}. When the energy $E$ lies in a gap, $\nu(E)$ is an integer, and it is the sum of the Chern numbers of all bands that lie below the given energy. We may rewrite this quantity in a form that is manifestly invariant under gauge transformations of Bloch vectors, and will be useful as we tackle the disordered case in Sec. \ref{k_disorder}. We define the projector $P_\mathbf{k} = P_\mathbf{k}(E)$ onto the Bloch states of energy below $E$. The projection operator is obtained from the Bloch eigenvectors of the tight-binding problem, and is therefore $\mathbf{k}$-dependent. We can rewrite Eq. (\ref{eq:k_nuE}) as
\begin{equation}
\label{eq:k_nuE_tr}
\nu(E) = \frac{1}{2\pi i}  \text{Tr}\left\{ P_\mathbf{k} \left[ \partial_{k_x} P_\mathbf{k}, \partial_{k_y} P_\mathbf{k} \right]  \right\}.
\end{equation}
\noindent The number $\nu(E)$ is an integer only when the energy lies between two bulk bands without touching them, inside of a gap, and it takes non-universal values for the slightest overlap in energy with the bulk bands (see Fig. \ref{fig:k_chern}). The quantization of this quantity corresponds to the existence of counter-propagating edge modes (see Sec. \ref{k_edgemodes}). The detuning of the phase $\phi$ will make $\nu(E)$ take a non-quantized value, which corresponds to edge modes that can decay into the bulk of the material. We corroborate that disorder is similar in effect to changes of the phase $\phi$: it also spreads the middle band leading to non-quantized values of the Chern number. This is explained in the next section and summarized in Fig. \ref{fig:k_chern}.

\subsection{Disordered system}
\label{k_disorder}

In this section we investigate the robustness of edge states in the disordered system. The Bloch state formulation used so far cannot be used since the system breaks translational invariance. To arrive at the formula for the Chern number in real-space for disordered systems, we start from the clean case. The trace in Eq. (\ref{eq:k_nuE_tr}) can be taken with respect to the basis of states localized at each site, $|\mathbf{m}\rangle$, where $\mathbf{m}=(m_1,m_2)$ indexes a site on the lattice. In this section we return to the $x,y$ basis, as this is more useful for deriving  formulae suitable for numerical computations. Letting as usual $\mathbf{r}_\mathbf{m} = m_1 \mathbf{\Delta}_1 + m_2 \mathbf{\Delta}_2$, 
\begin{eqnarray}
&&\nu(E) =  \nonumber \\
&&-\lim_{N \rightarrow \infty} \frac{2\pi i }{N} \sum_{\mathbf{m}} \langle \mathbf{r}_\mathbf{m} | P(E) \left[ -i [x, P(E)], -i[y, P(E)]  \right] | \mathbf{r}_\mathbf{m} \rangle,\nonumber \\
\;
\end{eqnarray}
\noindent where $P(E)$ is the Fourier transform of the operator $P_\mathbf{k}(E)$ defined above, and projects onto states with energy below $E$. The total number of sites in the system is denoted by $N$. In the thermodynamic limit for a clean system, this formula is exactly that from  Eqs. (\ref{eq:k_nuE}) and (\ref{eq:k_nuE_tr}). Its advantage is that it allows us to compute the Chern number in a disordered system. In Ref. \cite{Bellissard} it was shown that the Chern number $\nu(E)$ is an integer as long as the energy $E$ belongs to a gap in the spectrum. The integer value can change to a different one only if the energy $E$ crosses a region of extended states, such as traversing a bulk band. 

When the system is disordered and finite, the problem of averaging over  disorder configurations arises along with that of taking the thermodynamic limit. The Chern number can be defined as an ensemble average over disorder configurations \cite{Prodan,Bellissard}
\begin{equation}
\label{eq:k_nuE_r}
\nu(E) = \int d\mu(\delta) \nu_\delta(E),
\end{equation}
\noindent where the disorder configurations $\delta$ are distributed according to a measure $d\mu(\delta)$, and $\nu_\delta(E)$ is the value of the Chern number for the specific disorder configuration. One defines a Hamiltonian $\mathscr{H}_\delta$ for each disorder configuration $\delta$, and associates the projector $P_\delta (E)$ to it as in the clean case, which gives a formula for $\nu_\delta(E)$:
\begin{eqnarray}
\label{eq:k_nuOmE_r}
&&\nu_\delta(E) = \nonumber \\
&&-\frac{2\pi i }{N} \sum_{\mathbf{m}} \langle \mathbf{r}_\mathbf{m} | P_\delta (E) \left[ -i [x, P_\delta (E)], -i[y, P_\delta (E)]  \right] | \mathbf{r}_\mathbf{m} \rangle, \nonumber \\
&&\;
\end{eqnarray}
\noindent where now $x,y$ are the position operators in cartesian coordinates. This equation has the property that the average over an ensemble of disorder configurations can replace the thermodynamic limit of a single system. The technical details, and reformulations suitable for numerics are reserved for the Appendix \ref{ap:k_chern}.

We have computed the Chern number for the Kagome system with a flat middle band using the methods described in \cite{Prodan}  at $\phi = \frac{\pi}{6}$, implemented on a lattice of $24 \times 24$ sites, or 192 unit cells. At each point, the Chern number was computed for $40$ disorder configurations and averaged. A white noise disorder potential (uncorrelated from site to site) was sampled on the lattice. For a disorder amplitude $W$, a random number between $\left[-\frac{W}{2}, \frac{W}{2} \right]$ is produced. The two types of disorder that can appear on the lattice are scalar disorder, on the on-site frequencies, which we denote $W_{site}$, and vector disorder on the hopping phases, denoted $W_\phi$. 

Fig. \ref{fig:k_chern} shows our results for $W_{site}$ between $0$ and $1.5 |t|$ and $W_\phi$ between $0$ and $\frac{3\pi}{10}$. The originally flat middle band spreads with disorder; in view of Sec. \ref{k_ldos}, the broadening of the middle band is associated with states leaking out of the hexagonal plaquettes due to the detuning of the flux. Comparatively, the role of scalar or vector disorder potentials is the same. We are not addressing here the interesting question of comparing the two types of disorder.

\section{Anisotropies}
\label{k_sym}
Recently it has become possible to realize, shift, and merge Dirac cones in optical lattices \cite{Esslinger, Duan}. Anisotropies on the honeycomb lattice have been studied \cite{StephanWeiKaryn}. In the cQED realization of the Kagome lattice, anisotropies are inherent: the on-site energies $\hbar \omega_{A,B,C}$ and hopping amplitudes $t_{A,B,C}$ obtained after integrating passive circuit elements are generally distinct (this is presented in detail in Ref. \cite{Kagpho}). Small anisotropies leave the main features of the bandstructure intact. Below is a quantiative analysis of the effects of anisotropies.

Dirac cones are stable due to discrete symmetries \cite{Manes,Montambaux}. The isotropic Kagome lattice (isotropic hopping amplitudes and site potentials), is symmetric under inversions with respect to the center of a hexagonal plaquette, and under rotations by $2\pi/3$, up to the permutation of the sublattice indices (lattice geometry in Fig. \ref{fig:k_BS}). Additionally, for special values of the phase $\phi$ the system is invariant under time-reversal symmetry. These three symmetries fix the position of the Dirac cones at the corners of the Brillouin zone between two bands. Two more Dirac cones are merged at the $\Gamma$ point (where the dispersion of one of the bands is flat, and of the other quadratic). If the rotational symmetry is broken due to anisotropy, then the Dirac cones can be shifted in $\mathbf{k}$-space. For strong enough anisotropy, Dirac cones meet at time-reversal invariant points  in the Brillouin zone, where they can annihilate. Such points are the $\Gamma$ point and the $M$ points. In the example below, Dirac points annihilate at the $M$ points.

Using the notations introduced in Sec. \ref{k_tbmodel}, let us define $c_{1,2,12} \equiv \cos(\alpha_{1,2,12})$. The Hamiltonian including anisotropies is
\begin{widetext}
\begin{equation}
\label{eq:k_HkAn}
\mathscr{H}_{\mathbf{k}} = 
\left(\begin{array}{ccc}
\hbar \omega_A & 2t_{AB} \cos \alpha_1 & 2 t_{AC} \cos \alpha_{2}\\
2t_{BA} \cos \alpha_1 & \hbar \omega_B & 2t_{BC} \cos \alpha_{12}\\
2t_{CA} \cos \alpha_{2} & 2t_{CB} \cos \alpha_{12} & \hbar\omega_C
\end{array}\right).
\end{equation}

\noindent which leads to the following equation for the three energy levels:

\begin{eqnarray}
-E^3 + \hbar(\omega_A + \omega_B + \omega_C)E^2 -\left(\hbar^2(\omega_A \omega_B+\omega_B \omega_C+\omega_C \omega_A)-4( |t_{AB}|^2 c_1^2 +|t_{AC}|^2 c_{2}^2 + |t_{BC}|^2 c_{12}^2)\right)E \;& & \nonumber\\
- 4 \hbar(\omega_A c_{12}^2 |t_{BC}|^2 + \omega_B c_{2}^2 |t_{AC}|^2 + \omega_C c_1^2 |t_{AB}|^2) + 16|t_{AB}t_{AC}t_{BC}|c_1 c_2 c_{12}  \cos(3\phi)  +\hbar^3\omega_A \omega_B\omega_C   &=& 0.
\end{eqnarray}
\end{widetext}

\noindent Let us consider, for example, the case when the hopping integrals are isotropic, but there is an anisotropy between the on-site energies $\hbar \omega_{A,B,C}$ at the level of each triangular plaquette. The spectrum can become degenerate at the $M$ points defined by $k_x = 0$, $k_y=\pm \frac{2\pi}{a\sqrt{3}}$. One has $c_1 = 1$, $c_2 = 0$, and $c_{12} = 0$. Then
for all $\phi$, a degeneracy occurs if $\omega_C =  \left(\omega_A + \omega_B \pm \sqrt{ (\omega_A -\omega_B)^2 + \frac{16}{\hbar^2}|t_{AB}|^2}\right)/2$. At the other two pairs of $M$ points, $k_x = \pm\frac{ \pi}{a}$, $k_y = \pm \frac{\pi}{a\sqrt{3}}$ and $k_x = \pm\frac{ \pi}{a}$, $k_y = \mp \frac{\pi}{a\sqrt{3}}$ the degeneracy occurs if the cyclic permutations of the condition above hold. This procedure allows us to produce a degeneracy between the upper and middle bands or the lower and middle bands (although all three bands cannot be degerate at the same $\mathbf{k}$).  The values for $\omega_{A,B,C}$ necessary for this degeneracy are on the order of the hopping amplitude $|t|$. The touching cannot be lifted by varying the phase $\phi$. The degeneracy at the $M$ points is quadratic in at least one direction in $\mathbf{k}$-space, which is a feature of the fact that two Dirac cones with linear dispersion annihilate at the time-reversal invariant point (see, for example Ref. \cite{Manes}).  One can tune the anisotropy as to annihilate a pair of Dirac cones between the lower and middle band, for example. As soon as the Dirac cones have annihilated, the bands are no longer degerate and the respective gap no longer carries edge modes.

In an experiment it is plausible to consider the case of anisotropic phases $3\phi_{\textit{up}}$ and $3\phi_{\textit{down}}$ around the up-pointing and down-pointing triangular plaquettes, respectively. This arrangement maintains the requirement that there is zero flux per unit cell, and the effect will be to merely shift the Dirac cones due to the anisotropy. 

\section{Interactions}
\label{k_int}
In this section we address the stability of the topological phase in the presence of interaction effects. A host of methods have been used to show that the topological phases are stable even in the presence of interactions \cite{StephanKaryn,RaghuTMI}.  At a simple perturbative level, we show that interactions cannot change the effective free particle dispersion bandstructure: they cannot cause the bands to become degenerate and reopen the gap. The general argument goes as follows: if the system is in a topological phase with gaps between bulk bands, then it is enough to show that for weak enough interactions the shifts in the energy levels are smaller than the free particle spectrum gap size of order the hopping strength $|t|$. 

The following are interaction Hamiltonians which can be potentially implemented in arrays of resonators. The boson Hubbard model was discussed, for example, in an important paper by M. P. A. Fisher \textit{et al.} \cite{Fisher}. The Jaynes-Cummings Hamiltonian \cite{JC}, on the other hand, models an interaction between lattice photons  and quantum two-level systems situated at each site. Both the boson Hubbard model and the Jaynes-Cummings model exhibit a phase transition from a superfluid to a bosonic Mott insulator state. Both interactions are in principle realizable in a cQED experiment \cite{FazioReview, HouckTureciKoch}. For completeness, we add a brief analysis of fermionic interactions in Appendix \ref{ap:k_fermions}.

\subsection{Bose-Hubbard model}
We start with the unperturbed Hamiltonian, the one presented in Eq. (\ref{eq:k_Hk}),
\begin{equation}
\label{eq:k_ap_Hk}
\mathscr{H}=\sum_{{\bf k} \in \text{BZ}}\psi_{{\bf k}}^{\dagger}\mathscr{H}_{{\bf k}}\psi_{{\bf k}} = \sum_{\alpha,\beta,{\bf k} \in \text{BZ}} a^\dagger_{\alpha, \mathbf{k}} \mathscr{H}_\mathbf{k}^{\alpha\beta} a_{\beta,\mathbf{k}},
\end{equation}
\noindent where the Greek indices indicate the sublattice in the spinor  $\psi_{{\bf k}}^{\dagger}=\left(a_{A{\bf k}}^{\dagger},a_{B{\bf k}}^{\dagger},a_{C{\bf k}}^{\dagger}\right)$. The interaction Hamiltonian for the Bose-Hubbard model is quartic in the creation and annihilation operators
\begin{eqnarray}
&&\mathscr{H}_{BH} = \frac{U}{2} \sum_\mathbf{m} a^\dagger_\mathbf{m} a_\mathbf{m} ( a^\dagger_\mathbf{m} a_\mathbf{m}  - 1)= \nonumber \\
&&\sum_{\mathbf{k}}\sum_{\alpha = A,B,C}\left( \frac{U}{2 N} \sum_{\mathbf{k}_1, \mathbf{k}_2} a^\dagger_{\alpha,\mathbf{k}_1 - \mathbf{k}} a^\dagger_{\alpha,\mathbf{k_2}+\mathbf{k}} a_{\alpha,\mathbf{k}_2} a_{\alpha,\mathbf{k}_1} \right), \nonumber \\
&&\; 
\end{eqnarray}
where $N$ is the number of unit cells and we have used the Fourier transform on each sublattice $\alpha = A,B,C$ as $a_{\alpha,\mathbf{k}}=\frac{1}{\sqrt{N}}\sum_{\mathbf{m}\;\in\;\alpha} e^{-i \mathbf{k} \mathbf{R}_\mathbf{m}} a_\mathbf{m}$. We would like to describe the spectrum of elementary excitations above the ground state. The minimum of the spectrum is at the $\Gamma$ point: the ground state of the bosonic system has all photons condensed at $\mathbf{k} = 0$ in the lowest band (see Fig. \ref{fig:k_BS}). Let us denote the minimum single particle energy at the $\Gamma$ point by $E_0$.

In the following treatment we shall restrict to a subspace of constant particle number $n$, where $n \equiv N n_0$. We have denoted by $N$ the number of unit cells in the lattice, implying that there are $3N$ sites, and by $n_0$ the number of condensate particles per unit cell. The spectrum of elementary excitations in the superfluid phase is determined by using the Bogoliubov approximation \cite{LandauLifshitz}, 
\begin{equation}
\label{eq:k_bops}
a_{\alpha,\mathbf{k}} \equiv \sqrt{N n_0} \delta_\mathbf{k} + b_{\alpha,\mathbf{k}}.
\end{equation}
\noindent Eq. (\ref{eq:k_bops}) defines the excited state operators $b_{\alpha,\mathbf{k}\neq 0}$. The full Hamiltonian can be brought into the following form
\begin{eqnarray}
\label{eq:k_totalBH}
&&\mathscr{H}_t = \mathscr{H} + \mathscr{H}_{BH} = \nonumber \\
&& E_{\text{G}} + \sum_{\mathbf{k}\alpha\beta} \left(b^\dagger_{\alpha, \mathbf{k}} \left(\mathscr{H}_\mathbf{k}^{\alpha\beta} - E_0 \delta^{\alpha\beta} \right) b_{\beta,\mathbf{k}}   \right) +\nonumber \\
&&\frac{U n_0}{2} \sum_{\mathbf{k}\neq 0,\alpha} \left( 2 b^\dagger_{\alpha,\mathbf{k}} b_{\alpha,\mathbf{k}}  +  b_{\alpha,\mathbf{k}} b_{\alpha,-\mathbf{k}} + b^\dagger_{\alpha,\mathbf{k}}b^\dagger_{\alpha,-\mathbf{k}} \right),
\end{eqnarray}
\noindent where $E_\text{G}\equiv n E_0$ is the total ground state energy, and the remainder describes the excitation spectrum. We shall only focus on the excitation spectrum and drop the ground state energy from our notation. The Hamiltonian can be diagonalized by a Bogoliubov transformation \cite{LandauLifshitz} from the particle operators $b_{\alpha,\mathbf{k}}$ to a new set of quasiparticle operators $\tilde{b}_{n,\mathbf{k}}$, which annihilate a quasiparticle in the $n^{th}$ band. The details of the Bogoliubov transformation are reserved for Appendix \ref{ap:Bogoliubov}. The Hamiltonian of Eq. (\ref{eq:k_totalBH}) can be recast into a diagonal form from which one can extract the quasiparticle dispersions, which we denote by $\xi_n (\mathbf{k})$, 

\begin{figure}[t!]
\includegraphics[width=0.63\linewidth]{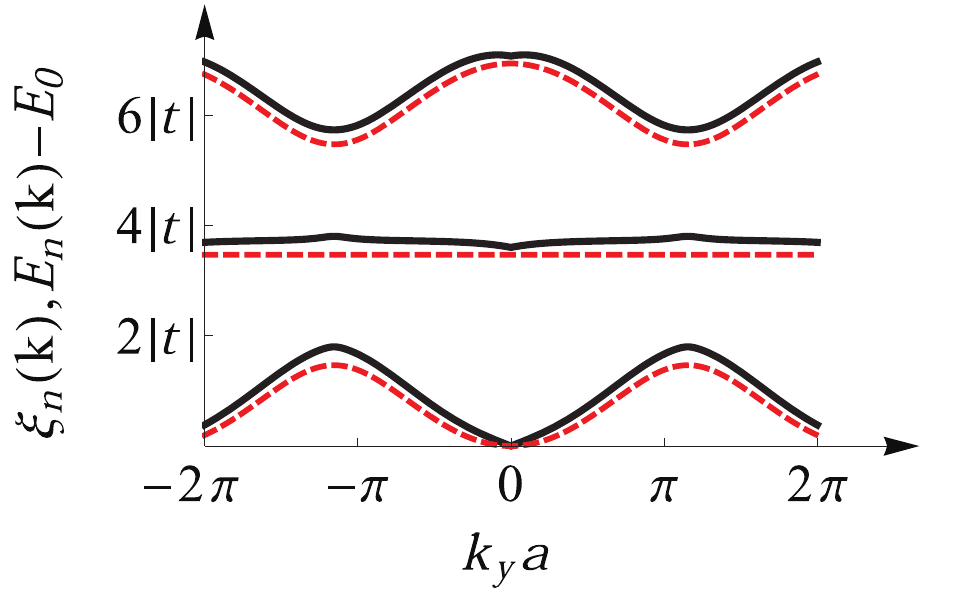}
\includegraphics[width=0.33\linewidth]{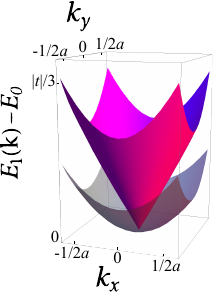}
\caption{\label{fig:k_BHresult} (Color online) The left panel shows the deviation of the effective quasiparticle dispersion $\xi_{n}(\mathbf{k})$ (black) from the free particle dispersion $E_n(\mathbf{k})$ (dashed, red) for the $\phi = \frac{\pi}{6}$ system with $U/|t| = 0.13$ (the plot is along the line $k_x=0$). Right panel: low energy quasiparticle spectrum exhibits linear dispersion (magenta) at the $\Gamma$ point $\mathbf{k} = 0$; plotted for comparison is the free particle dispersion (light gray).}
\end{figure}

\begin{equation}
\label{eq:k_xik}
\mathscr{H}_t = \sum_{\mathbf{k},n} \xi_n(\mathbf{k}) \tilde{b}_{n\mathbf{k}}^\dagger \tilde{b}_{n\mathbf{k}}.
\end{equation}

In the following, we discuss the stability of the topological phase when the Bose-Hubbard interaction is introduced. In general, if the interactions induce no band crossings, the Chern numbers on individual bands will be conserved, and the topological phase is maintained. The Hubbard interaction induced dispersion $\xi_n(\mathbf{k})$ can only change significantly from the free particle dispersion $E_n(\mathbf{k})$ of Eq. (\ref{eq:k_evals}) in the vicinity of the band minimum at the $\Gamma$ point, where the quasiparticle has a linear dispersion (the sound mode). Fig. \ref{fig:k_BHresult} shows the quasiparticle dispersion in comparison to the free particle dispersion, in the flat band system at $\phi=\frac{\pi}{6}$. As long as the Bose-Hubbard coupling $U$ is on the order of $|t|$ or smaller, the shift in the energy levels cannot make the gap close and reopen, and the topological phase is protected. In the weak interaction regime, the middle flat band can acquire a dispersion: This situation is similar to the effect of detuning the hopping phase $\phi$, or the effect of disorder. 

\subsection{Jaynes-Cummings interaction}

We proceed with the Jaynes-Cummings interaction \cite{JC} which couples photons to quantum two-level systems at each site. The Jaynes-Cummings system exhibits a quantum phase transition of polaritons from a superfluid phase to a Mott insulating phase \cite{JensKarynJC}, resembling that of the Bose-Hubbard model. 
The system is governed by the following Hamiltonian which describes the tight-binding photons, the two-level systems, and the coupling between these
\begin{eqnarray}
\label{eq:k_tbgeneric}
\mathscr{H}_t &=& \sum_{\mathbf{m},\mathbf{n}} t_{\mathbf{m},\mathbf{n}} a^\dagger_\mathbf{m} a_\mathbf{n}  + \sum_\mathbf{m} \epsilon \sigma_\mathbf{m}^+ \sigma_\mathbf{m}^- + \sum_\mathbf{m} g (\sigma_\mathbf{m}^+ a_\mathbf{m} + \text{h.c.})  \nonumber \\
&&\;
\end{eqnarray}
where we have written the tight-binding Hamiltonian in a more generic form as a sum over all pairs of sites, in terms of generic hopping integrals $t_{\mathbf{m},\mathbf{n}}$. Note that here the on-site energies $\hbar \omega$ have been included in the diagonal terms $t_{\mathbf{m},\mathbf{m}}$. Here $\sigma_\mathbf{m}^{\pm}\equiv \sigma_{x,\mathbf{m}} \pm i\sigma_{y,\mathbf{m}}$ are Pauli ladder operators describing a two-level system at lattice coordinate $\mathbf{m}$. 

To characterize the stability of the anomalous Hall phases in the presence of a weak (finite) Jaynes-Cummings interaction, in Appendix \ref{ap_photoneffec}, we find an effective low-energy Hamiltonian for the photon sector $\mathscr{H}^{\textit{eff}}$. We consider the following limit of small conversion $g$, 
\begin{equation}
g \ll  |t| < \hbar \omega.
\end{equation}
This assumption is prompted by the following stability condition specific to the Jaynes-Cummings lattice \cite{JensKarynJC}. If the ground state energy of the single particle spectrum, $E_0$, has a negative value, there is no mechanism to limit the number of photons that populate the ground state, and therefore the system becomes unstable. We therefore need $E_0 \geq 0$, which translates to a condition that $\hbar \omega > z|t|$, where here $z=4$ is the coordination number of the Kagome lattice. In a typical cQED experiment, we expect $|t|$ to not exceed $100$ MHz and $\omega$ to lie in the GHz range. Additionally, under reasonable experimental conditions the two level system excitation energy $\epsilon$ and the (renormalized) resonator frequencies $\omega$ will both lie in the (microwave) GHz range. The condition that $\hbar \omega \approx \epsilon$ is sufficient to order the two-level systems in the ground state, and the determination of an effective photon-photon interaction is well motivated. This, as we show below, leads to a positive value of the Hubbard coupling, i.e. a repulsive interaction.

To fourth order in the Jaynes-Cummings coupling $g$, we obtain the following effective low-energy Hamiltonian:
\begin{eqnarray}
\label{eq:effectiveQHQ}
&& \mathscr{H}^{\textit{eff}}  = \nonumber \\
&& E_0 +  \frac{f_2 g^2}{E_0 - \epsilon}  a^\dagger_{l 0} a_{l0} +     \frac{f_4 g^4}{( 2 E_0 - \epsilon )^3} a^\dagger_{l0} a_{l0} a_{l0}^\dagger a_{l0} ,
\end{eqnarray}
\noindent where $f_2$ and $f_4$ are positive dimensionless factors which can be found in Appendix \ref{ap_photoneffec} and the operator $a_{l0}^\dagger$ creates a particle in the ground state of the photon tight-binding Hamiltonian (lower band at the $\Gamma$ point, $\mathbf{k}=0$). Eq. (\ref{eq:effectiveQHQ}) shows that to lowest order the interaction shifts the (lowest) band minimum by an amount $\frac{f_2 g^2}{E_0-\epsilon}$. The next non-vanishing contribution appears at the quartic order. The repulsive character is obtainable under reasonable experimental conditions for the Jaynes-Cummings model, where both $\hbar\omega$ and $\epsilon$ would be in the microwave range. One way to understand the presence of a repulsive photon-photon interaction is by noting that the two-level systems are bosons with hard-core on-site repulsion. More precisely, if two-level systems are represented by bosons with interactions, the Jaynes-Cummings term can be absorbed by a shift of the bosonic two-level system field, and then the interaction term of the two-level systems also generates an on-site interaction for the photon in the shifted basis. The correspondence between the Jaynes-Cummings lattice and Bose-Hubbard model has also been demonstrated through a field theory approach close to the Mott-superfluid transition of light \cite{JensKarynJC}. In the calculation that led to Eq. (\ref{eq:effectiveQHQ}), we have considered the ground state with all photons at wavevector $\mathbf{k}=0$. This does not provide precise information about the range of the interaction, but it is telling that a Hubbard-type repulsive photon-photon interaction can emerge. Finally, as long as the perturbation $g$ is weak compared to the size of the gap, which is of order $|t|$, the detuning $\epsilon$, and the on-site energy $\hbar\omega$, the interactions cannot cause degeneracies, and the topological phase will be stable.

\section{Conclusions}
Motivated by the recent experimental progress in the context of arrays of electromagnetic superconducting resonators \cite{Houck}, we have investigated the anomalous Hall effect of light on the Kagome lattice with artificial gauge fields \cite{Kagpho}. The photonic system here exhibits equivalents of the quantum Hall effect without Landau levels, and the anomalous Hall effect with a non-quantized Chern number. In particular, we have shown that a topologically trivial band can affect the quantization of Chern numbers as well as the robustness of the chiral edge modes. We have discussed observables which are accessible experimentally. We have introduced a method to measure Berry's phases around loops of constant energy in the Brillouin zone. The method is based solely on wavepacket interference and can be used to determine band Chern numbers or the photonic equivalent of the anomalous Hall response. It provides an alternative to a recent method proposed to measure line integrals of the Berry gauge field in cold-atomic systems, which relies on the measurement of group velocities of wavepackets and a force-reversal protocol \cite{PriceCooper}. In cQED systems, this is realized by adjusting the frequency of each resonator individually, which has become experimentally possible \cite{Greentree2,Johansson, Sandberg,DelsingPRL}. It shall be noted that interference experiments can also be envisioned to probe the Landau levels, emerging when placing the Kagome lattice in a uniform magnetic field. An open and interesting question to investigate in the near future would be the influence of artificial gauge fields on the superfluid-Mott transition of light in cQED photon-based lattices, following a similar line of thought to that used in recent theoretical investigations  \cite{RaghuTMI, Peter}. Recent progress in this direction has been realized in Ref. \cite{Arun} where a chiral Mott insulator with a gap to all excitations and staggered fluxes has been found. Transport of microwave photons through cQED lattices with artificial gauge fields and disorder would be an interesting topic to explore. Another relevant subject to consider more thoroughly both theoretically and experimentally is the realization of quantum impurity models exhibiting fingerprints of many-body physics such as the Kondo model and Spin-Boson models \cite{KLH2, Shen, Longo, Baranger}.

We acknowledge useful discussions with Stephan Rachel, Peter Orth and Arun Paramekanti. AP and KLH are supported by the NSF through DMR-0803200 and DMR-0653377. AAH acknowledges support from the NSF Grant DMR-0953475.

\appendix
\section{Synthetic gauge fields}

In this appendix we present a more detailed derivation of the results in Sec. \ref{k_synthetic}. 

\subsection{Magnetic field}
\label{ap:synthetic}

This derivation starts from an idea used in Ref. \cite{Kolovsky} for square lattices. Consider the time-dependent Hamiltonian of Eq. (\ref{eq:k_Htau}), which we reproduce here:
\begin{equation}
\mathscr{H}_\tau = \mathscr{H} + \sum_\mathbf{m} \left(\hbar\omega + \hbar \omega_0 \cos \left( \Omega \tau + m_2 \theta  \right) m_1 \right) a^\dagger_\mathbf{m} a_\mathbf{m}.
\end{equation}
We construct the solution to the Schr\"odinger equation as follows. If the hopping is supressed, $|t|=0$, then the following function solves the time-dependent Schr\"odinger equation:
\begin{eqnarray}
\label{eq:k_phimtau}
|\psi \rangle &=& \sum_\mathbf{m} e^{i\phi_\mathbf{m}} |\mathbf{m}\rangle ,\nonumber \\
\phi_\mathbf{m}(\tau) &=& -\omega \tau - \frac{\omega_0}{\Omega} \sin\left(\Omega \tau + \theta m_2 \right)m_1. 
\label{eq:k_phim}
\end{eqnarray}
The solution for $|t|\neq 0$ is constructed from this as 
\begin{equation}
|\psi \rangle = \sum_\mathbf{m} d_\mathbf{m} e^{i\phi_\mathbf{m}} |\mathbf{m}\rangle,
\end{equation}
\noindent where the $d_\mathbf{m}$ must now obey the following differential equation
\begin{equation}
i\hbar \dot{d}_\mathbf{m} = \sum_{\mathbf{n}} t_{\mathbf{m},\mathbf{n}} e^{i\left( \phi_\mathbf{m} - \phi_\mathbf{n} \right)} d_\mathbf{n}.
\end{equation}
\noindent In this last equation, the $t_{\mathbf{m},\mathbf{n}}$ is the tight-binding hopping integral, which takes the value $|t|e^{\pm i \phi}$ for nearest-neighbors and zero otherwise, and $\phi_\mathbf{m}$ are the phases computed in Eq. (\ref{eq:k_phim}). The perturbation has induced a time-dependent phase factor $e^{i\left( \phi_\mathbf{m} - \phi_\mathbf{n} \right)}$ which we will now simplify by keeping only those parts that oscillate very slowly (rotating wave approximation). We shall use the following expansion
\begin{equation}
\label{eq:k_besselexpansion}
e^{i z \sin \alpha} = \sum_{l = - \infty }^{+\infty} e^{i l \alpha} \mathcal{J}_l (z),
\end{equation}
\noindent where $\mathcal{J}_l(z)$ are the Bessel functions of the first kind. Upon inspection of the expansion in Eq. (\ref{eq:k_besselexpansion}) and of the phase $\phi_\mathbf{m}$ in Eq. (\ref{eq:k_phimtau}) we find that, in general, one must have the driving frequency $\Omega$ be an integer multiple of the on-site frequency $\omega$, i.e. $l \omega$, in order to obtain at least one time-independent term in the expansion. The largest contribution is obtained if we take $\Omega=\omega$.  We obtain the following effective changes to the hopping amplitudes
\begin{eqnarray}
&& \text{bond} \parallel \hat{\mathbf{\Delta}}_1: \nonumber \\
&&\;\; |t| e^{-i \phi} \rightarrow |t| e^{-i \phi}\; \cdot\; \mathcal{J}_{-1}\left( \frac{\omega_0}{\Omega} \right) e^{-i{m_2 \theta}}, \nonumber \\
&& \text{bond} \parallel \hat{\mathbf{\Delta}}_2: \nonumber \\
&&\;\; |t| e^{-i \phi} \rightarrow |t| e^{-i \phi} \nonumber \\
&& \text{bond} \parallel \left(\hat{\mathbf{\Delta}}_2 - \hat{\mathbf{\Delta}}_1\right): \nonumber \\
&&\;\;|t| e^{-i \phi} \rightarrow |t| e^{-i \phi}\; \cdot\; \Big( \mathcal{J}_{-1} \left( \frac{\omega_0}{\Omega} (m_1+1) \right) e^{-i\theta m_2} + \nonumber \\
&& \;\;\;\;\;\;\;\;\;\;\;\;\;\;\;\;\;\;\;\;\;\;\;\;\;\;\;\;\;\;\;\;\;\;\;\; \mathcal{J}_{-1} \left(- \frac{\omega_0}{\Omega} m_1 \right) e^{-i\theta(m_2+1)}  \Big). \nonumber \\
\;
\end{eqnarray}
\noindent The ratio $\frac{\omega_0}{\Omega}$ provides an additional experimental parameter to tune the hopping strength via the Bessel functions of the first kind $\mathcal{J}_{-1}$.  

The time-dependent perturbation has induced spatially dependent phases and a dressing of the hopping integral $|t|$. These spatially dependent phases mimic the phases that would be produced by a gauge field in the minimal substitution. The fact that an additional phase is acquired along oblique bonds parallel to $\hat{\mathbf{\Delta}}_2 - \hat{\mathbf{\Delta}}_1$ implies that at the level of each unit cell the phases will correspond to a field that is non-uniform across the unit cell. However, the total phase acquired by a photon traversing around a parallelogram unit cell of area $|\mathbf{\Delta}_1 \times \mathbf{\Delta}_2|$, see Fig. \ref{fig:k_BS}, is going to be a constant equal to $f=2\theta$. Since the uniformity of the field at the level of the unit cell can be recovered by a gauge transformation, we perform all of our calculations (Appendix \ref{ap_butterfly}) for a uniform field. 

\subsection{Spectrum in a magnetic field}
\label{ap_butterfly}
In this appendix we show the detailed calculations for the spectrum of the Kagome system placed in a uniform magnetic field $\mathscr{B}_s$. The phase acquired by a particle along an elementary parallelogram unit cell of the Kagome lattice (see Fig. \ref{fig:k_BS}) is $f\equiv\frac{\mathscr{B}_s a^2 \sqrt{3}}{2}\equiv 8\pi f \equiv \sqrt{3} b a$, where $f$ is dimensionless, and $b$ has units of inverse length. The phase acquired by a photon around the unit cell is eight times the phase acquired on a counter-clockwise loop around a triangular plaquette. Let us for simplicity pick the following gauge field in the Landau gauge, $\mathscr{A}_s = ( -\mathscr{B}_s y, 0 )$, where the two components are cartesian. Due to the presence of a magnetic field, the Hamiltonian will couple $k_y$ to $k_y \pm \frac{b}{2}$ and to $k_y \pm b$. Without loss of generality, we take $\hbar \omega = 0$. The Hamiltonian reads

\begin{eqnarray}
\mathscr{H} &=& |t| e^{i \phi} \sum_\mathbf{k} e^{i \mathbf{k}.\mathbf{\Delta}_1} a^\dagger_{A k_x, k_y} a_{B k_x,k_y+b} + a^\dagger_{A k_x, k_y} a_{B k_x,k_y-b}  \nonumber \\
&&+ e^{i\frac{f}{16}} \Big( e^{-i\mathbf{k}.\mathbf{\Delta}_2} a^\dagger_{C,k_x,k_y} a_{B,k_x, k_y+\frac{b}{2}} + a^\dagger_{C,k_x,k_y} a_{B,k_x, k_y-\frac{b}{2}} \Big) \nonumber \\
&&+ e^{-i\frac{f}{16}} \Big( e^{-i\mathbf{k}.(\mathbf{\Delta}_1-\mathbf{\Delta}_2)-i \frac{b a \sqrt{3}}{8}} a^\dagger_{A,k_x,k_y} a_{C,k_x, k_y+\frac{b}{2}} \nonumber \\ 
&&+ e^{i \frac{ba \sqrt{3}}{8}} a^\dagger_{A,k_x,k_y} a_{C,k_x, k_y-\frac{b}{2}} \Big) + h.c.
\end{eqnarray}
For rational values of $f = \frac{p}{q}$, where $p,q$ are relatively prime integers, one can reduce the Brillouin zone from the original $\left[0,\frac{2\pi}{a}\right]\times \left[0,\frac{4\pi}{a\sqrt{3}}\right]$ to the magnetic Brillouin zone $\left[0,\frac{2\pi}{q a}\right]\times \left[0,\frac{4\pi}{a\sqrt{3}}\right]$. The couplings between different momenta disappear and we have replaced the original problem with that of a periodic one-dimensional chain of $3q$ sites. Let us take a generic wavefunction to be 
\begin{equation}
|\psi\rangle = \sum_{n=0,\alpha=A,B,C}^{q-1} \psi_{\alpha n} a^\dagger_{\alpha k_x,k^0_y+n\frac{b}{2}} |0\rangle,
\end{equation}
\noindent where $a^\dagger_{\alpha k_x, k^0_y + n\frac{b}{2}}$ creates a photon on sublattice $\alpha$ at a given momentum. Then the Schr\"{o}dinger equation is equivalent to the following set of three Harper equations
\begin{widetext}
\begin{eqnarray}
E_{k_x,k_y^0} \psi_{A m} &=& |t| e^{i\phi-i\frac{f}{16}} \Big( e^{+i\frac{b\sqrt{3}}{2}} \psi_{C,m-1} +e^{-i \left(k_x,k_y^0+m\frac{b}{2}\right).(\mathbf{\Delta}_1 - \mathbf{\Delta}_2) } \psi_{C,m+1}  \Big) +|t|e^{-i\phi}\Big( e^{-i\left( k_x, k_y^0 + m \frac{b}{2}\right).\mathbf{\Delta}_1} \psi_{B,m+2} + \psi_{B,m-2}\Big) \nonumber \\
E_{k_x,k_y^0}\psi_{Bm} &=& |t|e^{i\phi} \Big( e^{i\left(k_x,k_y^0+(m-2)\frac{b}{2}\right).\mathbf{\Delta}_1}\psi_{A,m-2}+ \psi_{A,m+2}\Big) +  |t|e^{-i\phi-\frac{f}{16}} \Big( e^{i\left(k_x,k_y^0+(m-1)\frac{b}{2}\right).\mathbf{\Delta}_2}\psi_{C,m-1}+ \psi_{C,m+1}\Big) \nonumber \\
E_{k_x,k_y^0}\psi_{Cm} &=& |t|e^{i\phi + i\frac{f}{16}} \Big( e^{-i\left(k_x,k_y^0+m\frac{b}{2}\right).\mathbf{\Delta}_2}\psi_{B,m+1}+ \psi_{B,m-1}\Big) \nonumber \\
&&\;\;\;\;\;\;\;\;\;\;\;\;\;\;\;\;+ |t|e^{-i\phi + i\frac{f}{16}}  \Big( e^{i\left(k_x,k_y^0+(m-1)\frac{b}{2}\right).(\mathbf{\Delta}_1-\mathbf{\Delta}_2)+i\frac{ba\sqrt{3}}{8}}\psi_{A,m-1}+ e^{-i \frac{ba \sqrt{3}}{8}}\psi_{A,m+1}\Big).\nonumber \\
\;
\end{eqnarray}
The solution to these equations gives the spectrum of the system for every rational flux $f=p/q$, and the pattern of splittings into magnetic subbands is known as the Hofstadter butterfly \cite{Hofstadter}. The spectrum of the problem is periodic in $f$ of period 2, for example $f=0$ and $f=2$ systems have the same spectrum etc.
\end{widetext}

\section{Evaluation of Chern numbers in disordered systems}
\label{ap:k_chern}
This appendix is dedicated to presenting more rigorous statements behind the real-space Chern number formulae of Eqs. (\ref{eq:k_nuE_r}) and (\ref{eq:k_nuOmE_r}). The general theory was introduced in \cite{Bellissard, Prodan}. To keep our discussion generic and more suitable for numerics, we shall rescale our Brillouin zone to $\left[0,\frac{2\pi}{a}\right]\times \left[0,\frac{2\pi}{a}\right]$. The results for the reciprocal unit cell of the Kagome lattice can be adapted from the following by the introduction of the appropriate Jacobian.

Consider a two-dimensional lattice with $K$ orbitals per site. We may take $\{|\mathbf{m}\alpha\rangle|\alpha=1,...,K\}$ to be kets localized at each site $\mathbf{m}=(m_1,m_2)$, corresponding to each orbital. This amounts to a basis of the full Hilbert space. The orbitals $\alpha$ may represent actual orbitals, or spin, isospin due to more ions per unit cell etc. The Bloch transformation is a unitary transformation that takes the Hilbert space $H$ to a direct sum of $\mathrm{C}^{K}$ spaces of $K$-tuples of complex numbers,  $U:H\rightarrow\oplus_{\mathbf{k}\in \text{BZ}}\mathrm{C}^{K}$.  A ket localized at site $\mathbf{m}$ transforms as
\begin{equation}
U|\mathbf{m} \alpha\rangle=\frac{1}{2\pi}\oplus_{\mathbf{k}\in \text{BZ}}e^{-i\mathbf{k}.\mathbf{r}_\mathbf{m}}\xi_{\alpha},\label{eq:Uaction}
\end{equation}
where $\mathbf{r}_\mathbf{m} = m_1 \Delta_1 + m_2 \Delta_2$, and $\xi_{\alpha}$ is the column vector of $K$ elements with the $\alpha^{th}$ entry equal to $1$, and the rest $0$. 

The Fourier transform of any operator $A$ is
\begin{equation}
U A U^{\dagger}=\oplus_{\mathbf{k}\in \text{BZ}}A(\mathbf{k}).
\end{equation}

The momentum space derivatives of an operator transform as
\begin{equation}
U^{\dagger}\left(\oplus_{\mathbf{k}\in \text{BZ}}\partial_{k_{j}}A(\mathbf{k})\right)U=-i\left[r_j,A\right],\;\; j=x,y,\label{eq:NonCommD}
\end{equation}


The following identity of traces has to hold for a clean infinite system
\begin{equation}
\frac{1}{(2\pi)^{2}}\int \text{Tr}\{A(\mathbf{k})\}d^{2}\mathbf{k}=\lim_{\mathcal{A}\rightarrow\infty}\frac{1}{\mathcal{A}}\text{Tr}_{\mathcal{A}}\{A\},\label{eq:NonCommInt}
\end{equation}
\noindent where $\text{Tr}$ is the trace over orbitals, whereas $\text{Tr}_{\mathcal{A}}$ is the trace over orbitals and over the sites included in a patch of the system of area $\mathcal{A}$.

Spatial disorder configurations $\delta$ are distributed according to a probability measure $d\mu(\delta)$, which by assumption obeys the following properties. The probability measure for disorder configurations is invariant under spatial translations (homogeneity); any subset of the disorder configuration space invariant under translations is a set of measure $0$. These properties amount to requiring that the probability measure for disorder configurations is ergodic with respect to spatial translations. Ergodicity implies that spatial and disorder configuration averages are interchangeable. The averaging in Eq. (\ref{eq:k_nuE_r}) can be performed over a small system with many disorder configurations to the same effect as on a system in the thermodynamic limit  with a single disorder configuration (self-averaging property).

For each disorder configuration $\delta$, one defines an operator $A_\delta$ (for example, the Hamiltonian $\mathscr{H}_\delta$). One cannot define $\mathbf{k}$-space calculus rules because the Fourier transform is no longer defined, but one may replace calculus rules in the Brillouin zone by the ``non-commutative'' rule
\begin{equation}
\partial_{k_j}A_{\delta} \rightarrow -i[r_{j},A_{\delta}],\;\forall\delta\label{eq:ncderivative}
\end{equation}

\noindent and integration becomes
\begin{equation}
\lim_{\mathcal{A}\rightarrow\infty}\frac{1}{\mathcal{A}}\text{Tr}_{\mathcal{A}}\{A\}=\int d\mu(\delta)\text{tr}_{0}\{A_{\delta}\},\label{eq:ncintegral}
\end{equation}
\noindent where the trace at the origin is given by $\text{tr}_{0}\{A\}\equiv \text{Tr}\{\pi_{0}A\pi_{0}\}$ involving the projector $\pi_{0}=\sum_{\alpha}|0,\alpha\rangle\langle0,\alpha|$ onto quantum states at the origin $\mathbf{m}=(0,0)$. 

With the help of such calculus rules, the Chern number in the disordered continuum becomes 
\begin{equation}
\nu(E)=2\pi i\int d\mu(\delta)\text{tr}_{0}\{P_{\delta}(E)[-i[x ,P_\delta(E)],-i[y,P_{\delta}(E)]]\}.\label{eq:CNProjectorDisordered}
\end{equation}
\noindent Ref. \cite{Bellissard} contains a proof of the fact that this expression for an infinite system is the analytical index of a Fredholm operator, and thus an \textit{integer}. This integer can only change if the energy $E$ crosses a region of extended states in the spectrum, i.e. something like a bulk band. For our numerical calculations, used the developments in Ref. \cite{Prodan}, which gives a fast converging formula for the Chern number of Eq. (\ref{eq:CNProjectorDisordered}). The Brillouin torus is discretized into $N_x \times N_y$ patches of momentum space area $\Delta_x \Delta_y$, and integration is approximated by Riemann summation. 

Derivatives in $\mathbf{k}$-space are approximated by a finite-differences formula
\begin{equation}
\partial_{k_{i}}P_{\mathbf{k}}\rightarrow\delta_{k_{i}}P_{\mathbf{k}_{\mathbf{n}}}\equiv\sum_{j=1}^{Q}c_{j}[P_{\mathbf{k_{n}}+j\Delta_{i}}-P_{\mathbf{k_{n}}-j\Delta_{i}}],\; i = x,y
\end{equation}
\noindent where the coefficients $c_m$ are chosen to insure exponential convergence in the limit of large $N_{x,y}$, and $Q$ is an adjustable number of order $N_{x,y}$. Then  the  Chern number becomes
\begin{eqnarray}
\nu(E)&=&\frac{1}{2\pi i}\sum_{k_{n}}\text{Tr}\{P_{k_{n}}[\delta_{k_{1}}P_{k_{n}},\delta_{k_{2}}P_{k_{n}}]\}\Delta^{2} \nonumber \\
&=&-2\pi i\text{Tr}\{\oplus_{k_{n}\in \text{BZ}}(P_{k_{n}}[\delta_{k_{1}}P_{k_{n}},\delta_{k_{2}}P_{k_{n}}])\}.\label{eq:ChernNumberDiscreteBZ}
\end{eqnarray}

We can now replace the derivatives in momentum space by their real-space counterparts through $U(e^{-i\mathbf{r}.\mathbf{\Delta}} P(E) e^{i\mathbf{r}.\mathbf{\Delta}}) U^{\dagger}=\oplus_{\mathbf{k} \in \text{BZ}} P_{\mathbf{k}+\mathbf{\Delta}}(E)$.  We arrive at the final formula for the Chern number
\begin{eqnarray}
&&\nu_\delta(E)= \nonumber \\
&&-\frac{2\pi i}{N_x N_y}\sum_{\mathbf{m},\alpha}\langle \mathbf{m}\alpha|P_{\delta}(E)[-i \lfloor x,P_{\delta}(E)\rfloor,-i\lfloor y,P_{\delta}(E)\rfloor]|\mathbf{m}\alpha\rangle,\nonumber \\
&&\;
\label{eq:DisorderedSystemDiscretizedChernNumber}
\end{eqnarray}
\noindent where the sum is now manifestly over a finite system. The bracket used above is derived from the original commutator $[x,P_\delta(E)]$
\begin{eqnarray}
&&\lfloor x,P_{\delta}(E)\rfloor = \nonumber \\
&&i\sum_{j=1}^{Q}c_{j}(e^{-i j x.\Delta_{x}} P_\delta(E) e^{i j x.\Delta_{x}}-e^{i j x.\Delta_{x}} P_\delta(E)e^{-i j x.\Delta_{x}}), \nonumber \\
&&\;
\label{eq:CleanLimitDiscretizedBracket}
\end{eqnarray}
\noindent and an analogous equation holds for $y$. The result needs to be averaged over disorder configurations,
\begin{equation}
\nu(E)=\int d\mu(\delta)\nu_{\delta}(E).
\end{equation}

\section{Interacting fermions on the Kagome lattice}
\label{ap:k_fermions}
In this appendix we analyse a Hamiltonian describing a fermionic nearest-neighbor interaction on the Kagome lattice:
\begin{equation}
\mathscr{H}_I = V \sum_{\langle ij\rangle} n_i n_j.
\end{equation}
The Fourier transformed interaction Hamiltonian reads
\begin{eqnarray}
\label{k_Hi}
\mathscr{H}_I &=& \frac{V}{N} \sum_{\mathbf{l}\mathbf{n}\mathbf{k}}\Big[ \cos\left(\mathbf{k}.\frac{\mathbf{\Delta}_1}{2} \right) a^\dagger_{A,\mathbf{l}-\mathbf{k}} a_{A,\mathbf{l}} a_{B,\mathbf{n}+\mathbf{k}}^\dagger a_\mathbf{B,n} + \nonumber \\
&&\cos\left(\mathbf{k}.\frac{\mathbf{\Delta}_2}{2} \right) a^\dagger_{A,\mathbf{l}-\mathbf{k}} a_{A,\mathbf{l}} a_{C,\mathbf{n}+\mathbf{k}}^\dagger a_\mathbf{C,n} + \nonumber \\
&&\cos\left(\mathbf{k}.\frac{\mathbf{\Delta}_1-\mathbf{\Delta}_2}{2} \right) a^\dagger_{C,\mathbf{l}-\mathbf{k}} a_{C,\mathbf{l}} a_{B,\mathbf{n}+\mathbf{k}}^\dagger a_\mathbf{B,n} \Big],
\end{eqnarray}
\noindent where $N$ is the number of unit cells on the lattice. We begin with the model containing a flat middle band at $\phi = \frac{\pi}{6}$.  Mean-field theory is performed easily if we switch to the band basis $(a^\dagger_{l\mathbf{k}},a^\dagger_{m\mathbf{k}},a^\dagger_{u\mathbf{k}})$, where the operators create particles of given momentum in each of the three bands, ``lower'', ``middle'', and ``upper''. In the band basis, the tight binding Hamiltonian $\mathscr{H}_\mathbf{k}$ of Eq. (\ref{eq:k_Hk}) describing free particles is diagonal,
\begin{equation}
\label{k_HkTdiag}
T_\mathbf{k}^\dagger \mathscr{H}_\mathbf{k} T_\mathbf{k}  = \text{diag}( E_l(\mathbf{k}), E_m(\mathbf{k}),  E_u(\mathbf{k}) ).
\end{equation}
\noindent The columns of $T$ are the Bloch eigenvectors of $\mathscr{H}_\mathbf{k}$. Then the two bases are related by the following unitary transformation,
\begin{eqnarray}
\label{k_tmatrix}
\left(\begin{array}{c} a_{A\mathbf{k}} \\ a_{B\mathbf{k}} \\ a_{C\mathbf{k}} \end{array} \right) = T_\mathbf{k} \left(\begin{array}{c} a_{l\mathbf{k}} \\ a_{m\mathbf{k}} \\ a_{u\mathbf{k}} \end{array} \right),\;\; \text{where} \nonumber \\
T_\mathbf{k} \equiv 
\left(\begin{array}{ccc} 
T_{Al}(\mathbf{k}) & T_{Am}(\mathbf{k}) & T_{Au}(\mathbf{k}) \\
T_{Bl}(\mathbf{k}) & T_{Bm}(\mathbf{k}) & T_{Bu}(\mathbf{k}) \\
T_{Cl}(\mathbf{k}) & T_{Cm}(\mathbf{k}) & T_{Cu}(\mathbf{k}) 
\end{array} \right).
\end{eqnarray}
At 1/3 filling, or when the lower band is completely occupied, the resulting mean-field Hamiltonian amounts to simply
\begin{equation}
\mathscr{H}_I ^{\text{MF}} = \frac{V}{3} \sum_{\mathbf{k} \in \text{BZ}} a^\dagger_{l\mathbf{k}} a_{l\mathbf{k}} + \textit{const},
\end{equation}
\noindent while at 2/3 filling, when both the lowest and the flat band are filled, the mean-field Hamiltonian takes the form
\begin{equation}
\mathscr{H}_I^\text{MF} = \frac{2 V}{3}\sum_{\mathbf{k} \in \text{BZ}} \left( a^\dagger_{l\mathbf{k}} a_{l\mathbf{k}}  + a^\dagger_{m\mathbf{k}} a_{m\mathbf{k}} \right).
\end{equation}
\noindent Then at mean-field level the nearest-neighbor interaction amounts to shifting the bands by an amount constant in $\mathbf{k}$-space. For strong enough interactions $V$, the bands will close gaps, but as long as $V<|t|$, which is the size of the band gap at $\phi = \frac{\pi}{6}$, the picture cannot be changed qualitatively and the topological phase is robust to this interaction effect. The cancelation of any $\mathbf{k}$-dependent shifts is a feature of the $C_3$ symmetry of the lattice.

\section{Bogoliubov transformation}
\label{ap:Bogoliubov}
In this appendix we present a way to diagonalize the Hamiltonian of Eq. (\ref{eq:k_totalBH}). The Hamiltonian matrix can be written in the following basis
\begin{equation}
\label{eq:k_phib}
\Phi^{\dagger}_\mathbf{k}  = \left( b_{A\mathbf{k}} b^\dagger_{A-\mathbf{k}} b_{B\mathbf{k}} b^\dagger_{B-\mathbf{k}} b_{C\mathbf{k}} b^\dagger_{C-\mathbf{k}}   \right),
\end{equation}
\noindent with the $b_{\alpha\mathbf{k}}$ operators as defined on each sublattice $\alpha \in \{A,B,C\}$ in Eq. (\ref{eq:k_bops}).
\begin{equation}
\mathscr{H}_t = \sum_{\mathbf{k} \in \text{BZ}} \Phi_\mathbf{k}^\dagger \mathscr{H}_{t\mathbf{k}} \Phi_\mathbf{k},
\end{equation}
where now the Hamiltonian may be expressed as
\begin{equation}
\label{eq:k_BogoH}
\mathscr{H}_{t\mathbf{k}}=
\left(\begin{array}{ccc}
h_{AA}(\mathbf{k}) & h_{AB}(\mathbf{k}) & h_{AC}(\mathbf{k}) \\
h_{BA}(\mathbf{k}) & h_{BB}(\mathbf{k}) & h_{BC}(\mathbf{k}) \\
h_{CA}(\mathbf{k}) & h_{CB}(\mathbf{k}) & h_{CC}(\mathbf{k}) 
\end{array}\right),
\end{equation}
\noindent whose elements are $2\times 2$ matrices defined as
\begin{eqnarray}
h_{\alpha\alpha}(\mathbf{k}) &=& 
\left(\begin{array}{ccc}
0 & \frac{U n_0}{2} \\
\frac{U n_0}{2} & \mathscr{H}^{\alpha\alpha}_{-\mathbf{k}} + \frac{U n_0}{2}
\end{array}\right), \nonumber \\
h_{\alpha\neq\beta}(\mathbf{k}) &=&
\left(\begin{array}{ccc}
\mathscr{H}^{\beta\alpha}_{\mathbf{k}} & 0 \\
0 & \mathscr{H}^{\alpha\beta}_{-\mathbf{k}}  
\end{array}\right),
\end{eqnarray}
where $\mathscr{H}_\mathbf{k}$ is the free particle Hamiltonian in Eq. (\ref{eq:k_Hk}). In the weak coupling limit $U \ll |t|$, the Bogoliubov transformation can be performed as follows. The $b$ operators can be transformed to quasiparticle operators $\tilde{b}$ via a canonical transformation
\begin{equation}
\Phi_\mathbf{k}  = B_\mathbf{k} \tilde{\Phi}_\mathbf{k},
\end{equation}
with $\tilde{\Phi}$ defined as in Eq. (\ref{eq:k_phib}) with $\tilde{b}$ replacing $b$  where the condition for $B_\mathbf{k}$ to preserve the bosonic commutation relations between the $\tilde{b}$ operators, $[\tilde{b}_{\alpha\mathbf{k}},\tilde{b}_{\beta\mathbf{k}'}^\dagger] = \delta_{\alpha\beta}\delta_{\mathbf{k}\mathbf{k}'}$ and $[\tilde{b}_{\alpha\mathbf{k}},\tilde{b}_{\beta\mathbf{k}'}] = 0$  is the following pseudo-unitarity condition
\begin{equation}
\label{eq:k_pseudounitary}
B_\mathbf{k}\Sigma B_\mathbf{k}^\dagger = \Sigma,\; \text{where}\; \Sigma = \text{id}_{3\times3} \otimes \sigma_3,
\end{equation}
\noindent where $\text{id}_{3\times3}$ is the identity acting on sublattice space and $\sigma_3$ is the third Pauli matrix. $B_\mathbf{k}$ must diagonalize the Bogoliubov Hamiltonian of Eq. (\ref{eq:k_BogoH}); let
\begin{equation}
B_\mathbf{k}^\dagger\mathscr{H}_{t\mathbf{k}} B_\mathbf{k} = \mathscr{K}_\mathbf{k}.
\end{equation}
From the condition in Eq. (\ref{eq:k_pseudounitary}) one can reexpress this as
\begin{equation}
\Sigma B_\mathbf{k}^\dagger \Sigma \Sigma \mathscr{H}_{t\mathbf{k}} B_\mathbf{k} = B_{\mathbf{k}}^{-1} \Sigma \mathscr{H}_{t\mathbf{k}}  B_\mathbf{k} = \Sigma \mathscr{K}_\mathbf{k}
\end{equation}
\noindent whence the matrices $\Sigma \mathscr{K}$ and $\Sigma \mathscr{H}_t $ are similar and it suffices to diagonalize the matrix $\Sigma \mathscr{H}_t$ to determine the spectrum of Bogoliubov quasiparticles.

\section{Effective photon Hamiltonian}
\label{ap_photoneffec}

In this appendix we derive an effective low-energy Hamiltonian $\mathscr{H}^{\textit{eff}}$ for the photons in the Jaynes-Cummings lattice model deep in the superfluid (delocalized) phase. 

The low energy state is the condensate state of $n$ photons and all of the $3N$ two-level systems are in their ground state. 
\begin{equation}
\label{eq:k_gndstJC}
|G_n\rangle =  |g_n \rangle \otimes \bigotimes_\mathbf{m} |\downarrow \rangle_\mathbf{m}
\end{equation}
\noindent where the operator $a_{l\mathbf{k}}^\dagger$ creates a photon in the lowest band at wavevector $\mathbf{k}$ and 
\begin{equation}
|g_n \rangle  = \frac{1}{\sqrt{n!}}\left(l^\dagger_{\mathbf{k}=0}\right)^n |0\rangle
\end{equation}
and $|0\rangle$ is the photon vacuum.

Consider the regime of weak coupling between photons and two-level systems described in the main text
\begin{equation}
g \ll |t| < \hbar \omega.
\end{equation}
Under realistic experimental conditions, $\hbar \omega \approx \epsilon$, which causes the two-level systems to be ordered in the ground state, and motivates our approach and to determine an effective photon-photon interaction starting with the ket in Eq. (\ref{eq:k_gndstJC}). Defining the projector on the ground state and the projector onto all excited states of the system as
\begin{equation}
Q = |G_n\rangle \langle G_n| \; \text{and}\; K = \text{id} - Q.
\end{equation}
These operators project into the subspace with $n$ polaritons, where $\sum_\mathbf{m} a^\dagger_\mathbf{m} a_\mathbf{m} + \sigma_\mathbf{m}^+ \sigma_\mathbf{m}^-$ is the polariton number. The indices $\mathbf{m}$ here represent the lattice coordinates.

Then
\begin{eqnarray}
\mathscr{H}_tQ &=& n E_l(\mathbf{k}=0)  Q + \nonumber \\
&& \sum_\mathbf{m} g a_\mathbf{m}|g_n\rangle \langle g_n|\otimes |\uparrow\rangle_\mathbf{m} \langle \downarrow |_\mathbf{m} \bigotimes_\mathbf{n} |\downarrow\rangle_\mathbf{n} \langle \downarrow |_\mathbf{n}, \nonumber \\
Q\mathscr{H}_t &=& n E_l(\mathbf{k}=0)  Q + \nonumber \\
&& \sum_\mathbf{m} g |g_n\rangle \langle g_n | a^\dagger_\mathbf{m} \otimes |\downarrow\rangle_\mathbf{m} \langle \uparrow |_\mathbf{m} \bigotimes_\mathbf{n} |\downarrow\rangle_\mathbf{n} \langle \downarrow |_\mathbf{n}. \nonumber \\
\end{eqnarray}
Above, $E_l(\mathbf{k})$ is the dispersion of the lowest band. Similarly we will denote by $E_{m,u}$ the dispersion relations for the middle and upper bands, respectively, of the original tight-binding Hamiltonian. The following operators are now completely specified using the equations above
\begin{eqnarray}
Q\mathscr{H}_tQ &=& n E_l(\mathbf{k}=0)  Q \nonumber \\ 
K\mathscr{H}_t Q &=& \mathscr{H}_t Q - Q\mathscr{H}_t Q \nonumber \\
Q\mathscr{H}_t K &=& Q \mathscr{H}_t - Q\mathscr{H}_t Q 
\end{eqnarray}
Finally the projections onto excited states can be expressed simply in terms of the projector onto the ground state
\begin{eqnarray}
K\mathscr{H}_t K &=& \mathscr{H}_t - Q\mathscr{H}_t - \mathscr{H}_t Q +Q\mathscr{H}_t Q \nonumber \\
&=& \mathscr{H}_p + \mathscr{H}_\epsilon + \mathscr{H}_g \nonumber \\
\mathscr{H}_p &=&  \sum_{\mathbf{m},\mathbf{n}} t_{\mathbf{m},\mathbf{n}} a^\dagger_\mathbf{m} a_\mathbf{n} - n E_l(\mathbf{k}=0) Q \nonumber \\
\mathscr{H}_\epsilon &=& \sum_\mathbf{m} \epsilon \sigma_\mathbf{m}^+ \sigma_\mathbf{m}^- \nonumber \\
\mathscr{H}_g &=& g\sum_\mathbf{m}\sigma^+_\mathbf{m} a_\mathbf{m} + \sigma^-_\mathbf{m} a^\dagger_\mathbf{m}  \nonumber \\
&& \;\;\;\;\;\;\; - Q\mathscr{H}_t - \mathscr{H}_t Q + 2Q\mathscr{H}_t Q. 
\end{eqnarray}
The effective low energy Hamiltonian is written as
\begin{equation}
\label{eq:effH}
\mathscr{H}^{eff} = Q \mathscr{H}_t Q + Q\mathscr{H}_t K \frac{1}{E - K \mathscr{H}_t K} K \mathscr{H}_t Q.
\end{equation}
A perturbation series in the small parameter $g$ can be obtained by expanding the denominator:
\begin{eqnarray}
\label{eq:expansion}
&&\frac{1}{E - K\mathscr{H}_t K} = \frac{1}{E - (\mathscr{H}_p + \mathscr{H}_\epsilon + \mathscr{H}}_g)  = \nonumber \\
&&\frac{1}{E -(\mathscr{H}_p + \mathscr{H}_\epsilon)} \frac{1}{\text{id}  - \frac{\mathscr{H}_g}{E -(\mathscr{H}_p + \mathscr{H}_\epsilon)}} = \nonumber \\
&&\frac{1}{E -(\mathscr{H}_p + \mathscr{H}_\epsilon)}  \cdot\Bigg[ \text{id} + \left( \frac{\mathscr{H}_g}{E -(\mathscr{H}_p + \mathscr{H}_\epsilon)} \right) \nonumber \\
&&+ \left( \frac{\mathscr{H}_g}{E -(\mathscr{H}_p + \mathscr{H}_\epsilon)} \right)^2 + O(g^3) \Bigg] \nonumber \\
\;
\end{eqnarray}
The main step is to invert
\begin{equation}
\frac{1}{E -(\mathscr{H}_p + \mathscr{H}_\epsilon)}
\end{equation}
\noindent where  
\begin{eqnarray}
\mathscr{H}_\epsilon = \sum_\mathbf{m} \epsilon \sigma_\mathbf{m}^+ \sigma_\mathbf{m}^-.
\end{eqnarray}
There is a basis in which the operator $\mathscr{H}_p + \mathscr{H}_\epsilon$ is diagonal:
\begin{eqnarray}
&&\mathscr{H}_p + \mathscr{H}_\epsilon = \nonumber \\
&& \sum_\mathbf{k} a^\dagger_{l\mathbf{k}} a_{l\mathbf{k}} E_l(\mathbf{k}) + a^\dagger_{m\mathbf{k}} a_{m\mathbf{k}} E_m(\mathbf{k}) + a^\dagger_{u\mathbf{k}} a_{u\mathbf{k}} E_u(\mathbf{k})  \nonumber \\
&&- n E_l(\mathbf{k} = 0) Q + \epsilon\sum_\mathbf{m} \sigma_\mathbf{m}^+\sigma_\mathbf{m}^-,
\end{eqnarray}
where we have introduced the operators creating a particle in the lower, middle and upper band as $a^\dagger_{l\mathbf{k}},a^\dagger_{m\mathbf{k}}$ and $a^\dagger_{u\mathbf{k}}$, respectively. These band basis operators are related to the original operators $a^\dagger_{A\mathbf{k}}$ etc. by a linear transformation which is written explicitly in Appendix \ref{ap:k_fermions}. The corresponding band dispersion relations are $E_{l,m,u}(\mathbf{k})$, and this form of the tight-binding Hamiltonian is simply the diagonal form of Eq. (\ref{eq:k_Hk}). We have now obtained a sum of decoupled Hamiltonians; the tight-binding part is already diagonal in the band basis introduced here; the two-level system piece is already diagonal in coordinate space. 

Let us compute the effective Hamiltonian up to lowest order contribution in $g$, from Eq. (\ref{eq:expansion}):
\begin{equation}
\label{eq:lowest}
\mathscr{H}^{eff} \supset Q \mathscr{H}_t Q + Q\mathscr{H}_t K \frac{1}{E -(\mathscr{H}_p + \mathscr{H}_\epsilon)} K \mathscr{H}_t Q.
\end{equation}
\noindent Only the terms containing $a_{l\mathbf{k}=0}$ contribute due to the projectors $Q$ at the two ends of the second term.  Then we only need to take the following terms for the evaluation of Eq. (\ref{eq:lowest}):
\begin{equation}
\mathscr{H}_p + \mathscr{H}_\epsilon =  a^\dagger_{l0} a_{l0} E_l(0)  - \epsilon\sum_\mathbf{m} \sigma_\mathbf{m}^+\sigma_\mathbf{m}^-.
\end{equation}
We have cast the denominator into a form which is diagonal with respect to both photon quantum numbers and two-level system quantum numbers. The $|\downarrow\rangle \langle \uparrow | $ of $Q\mathscr{H} K$ and the $|\uparrow\rangle \langle \downarrow | $ of $K\mathscr{H} Q$ acting on the diagonal matrix in between them will yield the projector onto the down state $|\downarrow\rangle \langle \downarrow |$, i.e. the second term in Eq. (\ref{eq:lowest}) will be a c-number times $Q$. More explicitly,
\begin{eqnarray}
&& Q\mathscr{H} K \frac{1}{E - a^\dagger_{l0} a_{l0} E_l(0) + n E_l(0) Q - \epsilon\sum_\mathbf{m} \sigma_\mathbf{m}^+\sigma_\mathbf{m}^-} K \mathscr{H} Q = \nonumber \\
&&= |g_n\rangle \langle g_n | \frac{g^2 \sum_\mathbf{m} a^\dagger_\mathbf{m}a_\mathbf{m}}{E - a^\dagger_{l0} a_{l0} E_l(0) - \epsilon }  |g_n\rangle \langle g_n | \otimes \bigotimes_\mathbf{m} |\downarrow\rangle _\mathbf{m} \langle \downarrow |_\mathbf{m} \nonumber \\
&&=\frac{n g^2}{E - (n-1)E_l(0)-\epsilon} Q = \frac{g^2 a^\dagger_{l0} a_{l0}  }{E - (n-1)E_l(0)-\epsilon} Q.
\end{eqnarray}
Then to lowest order $g^2$,
\begin{equation}
\mathscr{H}^{\textit{eff}} = Q \left( E_l(0) + \frac{g^2}{E - (n-1)E_l(0)-\epsilon} \right) a^\dagger_{l0} a_{l0}  Q
\end{equation}
To retrieve the effective Hamiltonian for the theory, we need to take the limit $E \rightarrow n E_l (\mathbf{k} = 0)$, i.e. this is only a valid expression close to the state ground state onto which $Q$ projects. Taking the limit above gives us an effective shift of the $\mathbf{k}=0$ energy. Peeling off the projector $Q$, we obtain the following effective low-energy Hamiltonian:
\begin{equation}
\mathscr{H}^{\textit{eff}} = \left( E_l(0) + \frac{g^2}{E_l(0)-\epsilon} \right) a^\dagger_{l0} a_{l0},
\end{equation}
which shows that at lowest order the interaction shifts band minimum by an amount $\frac{g^2}{E_l(0)-\epsilon}$. 

Higher order contributions yield effective interactions for the photons. The next nonvanishing contribution is in $g^4$, and comes from the term
\begin{eqnarray}
\mathscr{H}^{\textit{eff}} &\supset& Q \mathscr{H}_t K \left( \frac{1}{ E - (\mathscr{H}_p + \mathscr{H}_\epsilon)} \right)^2 \mathscr{H}_g \cdot \nonumber \\
&& \cdot \left( \frac{1}{ E - (\mathscr{H}_p + \mathscr{H}_\epsilon)} \right) \mathscr{H}_g  K \mathscr{H}_t Q 
\end{eqnarray}
After a similar calculation to the one above, we obtain the following form for this term, 
\begin{equation}
\mathscr{H}^{\textit{eff}} \supset \left(\frac{g^4}{( 2 E_l(0) - \epsilon )^3 } \sum_{\mathbf{m},\mathbf{m}'} a^\dagger_{\mathbf{m}} a^\dagger_{\mathbf{m}'} a_{\mathbf{m}} a_{\mathbf{m}'}   \right).
\end{equation}
In the main text, we have denoted the ground state energy of the single particle tight-binding model as $E_l (0) = E_0$. Peeling off the projector onto the ground state, $Q$, we may rewrite the effective Hamiltonian $\mathscr{H}^{\textit{eff}}$ as
\begin{eqnarray} 
E_0 +  \frac{f_2 g^2}{E_0 - \epsilon}  a^\dagger_{l 0} a_{l0} +     \frac{f_4 g^4}{( 2 E_0 - \epsilon )^3} a^\dagger_{l0} a_{l0} a_{l0}^\dagger a_{l0},
\end{eqnarray}
where $f_2$ and $f_4$ are dimensionless quantities defined as $f_2 \equiv ( |T_{Al 0}|^2 + |T_{Bl0}|^2 + |T_{Cl0}|^2 )$ and $f_4 \equiv (|T_{Al 0}|^4 + |T_{Bl0}|^4 + |T_{Cl0}|^4)$, where the $T$'s are the elements of the unitary transformation from sublattice basis to band basis, introduced in Appendix \ref{ap:k_fermions}: $a_{\alpha\mathbf{k}} = T_{\alpha l\mathbf{k}} a_{l\mathbf{k}} + T_{\alpha m \mathbf{k}} a^\dagger_{m\mathbf{k}} + T_{\alpha u\mathbf{k}} a^\dagger_{u\mathbf{k}}$ for each $\alpha = A,B,C$.

Under realistic experimental conditions, where both $E_0$ and $\epsilon$ are in the microwave range, one can obtain a repulsive interaction with $g^4/(2E_0 - \epsilon)^3 > 0$. Since we have only considered the ground state in which all photons are at wavevector $\mathbf{k}=0$, this approach does not provide precise information about the range of the interaction, but it is telling that a Hubbard-type repulsive photon-photon interaction can emerge.

\end{document}